\documentclass[a4paper,12pt]{article}
\pdfoutput=1 

\usepackage{jheppub} 

\usepackage[T1]{fontenc} 
\usepackage{amsmath,amssymb}
\usepackage{dsfont}
\usepackage{upgreek}
\usepackage{slashed}
\usepackage{appendix}

\newcommand{\half}{{{\textstyle\frac{1}{2}}}}
\newcommand{\quarter}{{{\textstyle\frac{1}{4}}}}
\newcommand{\be}{\begin{equation}}
\newcommand{\ee}{\end{equation} }
\newcommand{\beqa}{\begin{eqnarray} }
\newcommand{\eeqa}{\end{eqnarray} }
\newcommand{\ba}{\begin{array}}
\newcommand{\ea}{\end{array}}
\newcommand{\nn}{\nonumber}

\newcommand{\R}{\ell}
\newcommand{\Rp}{\ell_{\scriptstyle{P}}}

\newcommand\Tr{{\rm Tr}}

\newcommand\cD{{\cal D}}

\newcommand\cF{{\cal F}}

\newcommand\cL{{\cal L}}
\newcommand\cM{{\cal M}}
\newcommand\cN{{\cal N}}

\newcommand\cQ{{\cal Q}}

\newcommand\cV{{\cal V}}

\def\l{\left}
\def\r{\right}

\preprint{KIAS-P14053}

\title{\bf{Supersymmetric Localization for BPS Black Hole Entropy: 1-loop Partition Function from Vector Multiplets}}
 \author[a]{\small{Rajesh Kumar Gupta},}
\author[b]{\small{Yuto Ito},}
\author[b]{\small{Imtak Jeon},}
\affiliation[a]{\small{ICTP, Strada Costiera 11, 34151 Trieste, Italy}}
\affiliation[b]{\small{School of Physics, Korea Institute for Advanced Study, Seoul 130-722, Korea}}
\emailAdd{rgupta@ictp.it}
\emailAdd{yito@kias.re.kr}
\emailAdd{imtakjeon@kias.re.kr}

\abstract{We use the techniques of supersymmetric localization to compute the BPS black hole entropy in $\mathcal N=2$ supergravity.  We focus on the $n_v+1$ vector multiplets on the black hole near horizon background which is AdS$_2 \times$ S$^{2}$ space.  We find the localizing saddle point of the vector multiplets by solving the localization equations, and compute the exact one-loop partition function on the  saddle point. Furthermore, we propose the appropriate functional integration measure. Through this measure, the one-loop determinant is written in terms of the radius of the physical metric, which depends on the localizing saddle point value of the vector multiplets. The result for the one-loop determinant is consistent with the logarithmic corrections to the BPS black hole entropy from vector multiplets.}

\begin{document} 
\maketitle
\flushbottom
\section{Introduction} 
A consistent theory of quantum gravity should be able to provide the statistical interpretation of Bekenstein-Hawking entropy which is given by one quarter of the area of the horizon in Planck units \cite{Bekenstein:1973ur,Hawking:1974sw}. String theory being a candidate for the quantum theory of gravity provides a natural framework to study classical and quantum properties of black holes. In last decade there has been tremendous progress in this direction in the large cases of supersymmetric extremal black hole after the work of Strominger and Vafa \cite{Strominger:1996sh}. In particular, now we have a very good understanding of statistical degeneracy for a large class of supersymmetric extremal black hole in $\mathcal N=4$ and $\mathcal N=8$ string theory which in the thermodynamic limit reduces to Bekenstein-Hawking entropy \cite{Dijkgraaf:1996it,Maldacena:1999bp,Lopes Cardoso:2004xf,Shih:2005uc,Shih:2005qf,Jatkar:2005bh,David:2006ji,Dabholkar:2006xa,David:2006yn,David:2006ru,Sen:2008ta}. In order to extend this comparison beyond thermodynamic limit, one needs to understand how to compute the corrections to Bekenstein-Hawking entropy in both microscopic and macroscopic level. In a quantum theory one would expect that both the microscopic and macroscopic entropy will receive corrections from perturbative and non-perturbative effects.  At the microscopic level understanding, these corrections involves computation of degeneracy to a greater accuracy and its asymptotic expansion \cite{Banerjee:2008ky,Mandal:2010cj}. On the other hand at the macroscopic level, one needs a full quantum generalization of the entropy formula. 

The area law is generalized to Wald entropy formula \cite{Wald:1993nt,Iyer:1994ys} to take into account the higher order derivative corrections which include the $\alpha'$-corrections in string theories. For the single centered extremal black case, the formula was further generalized by Sen \cite{Sen:2008yk,Sen:2008vm}  based on AdS$_2/$CFT$_1$. The extremal black hole has the AdS$_2$ factor in its near horizon geometry, so it is of the from, AdS$_2\times K$, in 4 spacetime dimensions. Here, $K$ becomes S$^2$ for the supersymmetric case because the supersymmetry requires the extremal black holes to be spherically symmetric. 
 According to this proposal, the full quantum entropy associated with the horizon of an extremal black hole is given in terms of expectation value of Wilson loop at the boundary of the AdS$_2$. The proposal takes the form,
\begin{equation}\label{QuantumS}
W(p,q)=\left<exp\left[-iq_i\oint d\theta A^i_\theta\right]\right>^{finite}_{AdS_2},
\end{equation}
where $<\, >^{finite}_{AdS_2}$ denotes the finite part of unnormalized Euclidean path integral and the quantum entropy associated with the horizon is given by
\begin{equation}
S_{hor}(p,q)=\ln W(p,q).
\end{equation}
Since the proposal involves the path integral over all fields including the metric, there is no notion of fixed background. But, as is denoted by the subscript in (\ref{QuantumS}), the boundary condition is fixed by the attractor values of the black hole background, which is the AdS$_2$ geometry. The Wilson loop wraps the boundary of AdS$_2$.  The insertion of the Wilson line at the boundary means that we change the boundary condition from Dirichlet to Neumann condition for gauge field. Neumann condition fixes the electric fields at the boundary i.e. electric charges and hence the proposal computes the entropy in the microcanonical ensemble. Further, we need to extract finite part of the functional integral to see the physically meaningful quantity.  Since there is an IR divergence due to the infinite volume of AdS$_2$ space, the IR divergence should be removed by regularization and the holographic renormalization.

The classical limit of this partition function reduces to exponential of Wald entropy. Furthermore, one can use this proposal to compute the full quantum corrections. It includes not only $\alpha'$ correction but also  $g_s$ quantum correction as well as the non-perturbative correction to the entropy. To compute this, one has to integrate over all string fields on each saddle point. Since this integral over all string fields is quite difficult and challenging,  the strategy we follow is to first integrate out all massive KK modes and stringy modes, and  write down a Wilsonian effective action. This effective action will be given in terms of few massless supergravity fields and include all higher derivative corrections 
 together with non-perturbative corrections coming from worldsheet instantons. Thus we are left with the path integral over massless fields with the above boundary conditions and we takes this as our starting point.
By computing the path integration,
the proposal of the quantum entropy function has been tested. Perturbative calculation on a classical saddle point and comparing it with the similar expansion on the microscopic side has led to perfect match of logarithmic correction in case of BPS black hole in $\mathcal N=4$ and $\cN=8$ supergravities \cite{Banerjee:2010qc,Banerjee:2011jp,Gupta:2013sva,Gupta:2014hxa} in $4$-dimensions and BMPV black hole in $5$-dimensions \cite{Sen:2012cj}.

The computation of the path integral can also be  performed by using,  so called, supersymmetric localization. 
It is a powerful method, making the exact  computation possible in a supersymmetric theory. 
This method has been used quite successfully in the cases of supersymmteric gauge theories in various dimensions and on various compact manifolds \cite{Pestun:2007rz,Kapustin:2009kz,Hama:2010av,Benini:2012ui,Doroud:2012xw,Hama:2012bg}. The argument of the localization principle is so general that this principle can also be applied to the supergravity computation. 
 The argument of the supersymmetric localization is following \cite{Schwarz:1995dg}.  Let us suppose that $\cQ$ be a fermonic symmetry which gives rise to a compact bosonic symmetry,
\be
\cQ^2= H\,.
\label{H}\ee
We would like to compute an integral of some $\cQ$ invariant function $h$ and $\cQ$ invariant action $S$,
\be
Z=\int {\rm d}\mu\, h\, e^{-S}\,,
\ee
where we let the measure ${\rm d}\mu$ is also invariant under the $\cQ$. We deform a partition function by adding the $\cQ$-exact function $\cQ V$ with parameter $t$,
\be
Z_{t}=\int d\mu\,h\,e^{-S-t \cQ V}\,.
\label{Zt}\ee
where $V$ is a fermionic function and invariant under the $H$-transformation.
Since the action $S$, measure $d\mu$ and the localization action are invariant under the the supersymmetry $\cQ$, the modified partition function $Z_t$ is independent of the parameter $t$.
\be
\frac{d}{d t} Z_t = -\int {\rm d}\mu\, \cQ V \, h \, e^{-S-t \cQ V}=-\int {\rm d}\mu\, \cQ\left(V \, h \, e^{-S-t \cQ V}\right)=0\,.
\ee
 In the limit $t \rightarrow \infty$, the semiclassical approximation with respect to $1/t$ is exact. 
  One left with the integration over the submanifold $\cM_\cQ$
\be
Z= Z_\infty= \int_{\cM_{\cQ}} d\mu_\cQ \,e^{-S}Z_{1-\text{loop}}\,,
\ee
where $\cM_\cQ$ is the manifold where $\cQ V=0$ and $d\mu_\cQ$ is the induced measure on the submanifold $\cM_\cQ$.  
 For supergravity case,  a rigid supersymmetry parameter can be chosen, where the $\cQ^2$ should preserve the asymptotic boundary conditions. In the case of the black hole entropy, we choose a Killing spinor of AdS$_2\times$S$^2$.

 
The supersymmetric localization principle requires the off-shell closure of the supersymmetry algebra.  The $\mathcal N=2$ supergravities coupled to vector multiplets in 4-dimensions has an off-shell formulation in terms of conformal supergravity \cite{deRoo:1980mm, de Wit:1980tn, de Wit:1984px}. It is a gauge theory, where all the $\cN=2$ superconformal symmetries are promoted to the local symmetries, which couples to the matter fields, and gauge equivalent to the Poincare supergravities. The Weyl multiplet  having off-shell degrees of freedom includes the gauge fields for all the local symmetries, where the graviton and gravitini are contained.   To have the degrees of freedom for $\cN=2$ Poincare supergravity, one needs to add additional matter multiplets which is called compensating multiplets.  One of the advantages of this formulation is that the off-shell supersymmetry algebra does not depend on the choice of prepotential and as a result the solution for the localization equations and the computation of one-loop partition function do not depend on the details of prepotential. 

To utilize the advantage of  the conformal supergravity, we use the freedom of a choice of the gauge condition. Note that the metic $g_{\mu\nu}$ in Weyl multiplet is not the physical metric and conformaly related to the metric in Einstein frame $G_{\mu\nu}$,
\be
g_{\mu\nu}=G_{\mu\nu}e^{K(X,\bar{X})}\,,
\ee
where $K(X,\bar{X})$ is the K$\ddot{\text a}$hler potential that is function of the scalars in the vector multiplets. A conventional gauge for the scale symmetry is choosing the $e^{K}=1$, and it constrains the  the degree of freedom of $n_v +1$ scalars.  Instead of this gauge, we use another choice: the radius $\R$ of the AdS$_2\times$S$^2$ metric $g_{\mu\nu}$ to be constant, and all the $n_v +1$ scalars to be free to fluctuate. Throughout this paper, we follow this gauge choice. Note that the conformal mode of the physical metric $G_{\mu\nu}$ is encoded in the fluctuating scalars in vectormultiplets.  

The application of the supersymmetric localization to quantum entropy function was initiated in \cite{Dabholkar:2010uh,Dabholkar:2011ec,Banerjee:2009af}.
 In the work of \cite{Dabholkar:2011ec}, the authors consider $\frac{1}{8}$th BPS black hole in $\mathcal N=8$ supersymmetric string theory for which microscopic answer is known. After considering truncations of $\mathcal N=8$ supergravity to $\mathcal N=2$ supergravity with only vector multiplets, and assuming that the one-loop determinant coming from localizing action is trivial, they find that the on-shell action evaluated on the localization solutions together with proper integration measure itself reproduces the modified Bessel function, which is the microscopic answer for $\frac{1}{8}$th BPS black hole in $\mathcal N=8$ theory. The agreement with microscopic answer is remarkable, however we still need to understand the assumptions taken in this process. 
The integration measure should be the result of the one-loop determinant coming from all the multiplets including Weyl multiplets and gravitini multiplets. 

It is the purpose of this paper to verify these assumptions. As a first step, we focus on the fluctuations of $n_v+1$ abelian vector multiplets and compute the $Z_{1-\text{loop}}$ while keeping the Weyl multiplet and all other multiplets to their classical near horizon background.
It is essentially equivalent to that we are considering fluctuation of vectormultiplets on the localizing saddle point of the Weyl multiplet as it is known that the Weyl multiplet localized to its on-shell background AdS$_2\times$S$^2$ \cite{Gupta:2012cy}.
In the computation of the functional integral, the analytic continuation could be a subtle issue because the Euclidean action is not positive definite. We will address two possible choices. One is motivated from the work of Pestun, Hama, Hosomich \cite{Pestun:2007rz,Hama:2012bg}, the other is from the work of Dabholkar \textit{et al.} \cite{Dabholkar:2010uh,Dabholkar:2011ec,Gupta:2012cy}. Although we will choose the former one throughout this paper as it seems conceptually easier and safer,  we will argue that both choices will be consistent. The definition of the functional integration measure would also be subtle. A non-linear sigma model specifies its non-trivial functional integration measure by the principle of ultra locality \cite{Polchinski:1985zf,Moore:1985ix}. We will follow this idea to suggest the path integration measure of the supereravities.


We summarize our results here. We first find the solutions of the localization equations using our choice of reality properties and find that the solutions of localization equations are labelled by 2 real parameters for each vector multiplet. 
We then compute the determinant of the quadratic fluctuations of the $\cQ$-exact deformations about the localization solution. Since we are dealing with the abelian vector multiplets, the answer does not seem to depend on the parameters of the localization solutions.  Also, since the off-shell supersymmetry transformations for the fields involve unphysical metric which has dilatation weight $-2$, our answer of the one-loop determinant seems not scale invariant if the ordinary path integration measure assumed. However, given that our calculation is in conformal supergravity where all the symmetries are realized as gauge symmetry, one would expect that with the gauge invariant measure the one-loop determinant should be scale invariant. We propose the scale invariant path integral measure involving vector multiplet fields including ghost fields. With the proposed measure we find that the answer does depend on the localization solution through the physical metric which is scale invariant. It produces the vector multiplet contribution to the classical measure assumed in \cite{Dabholkar:2010uh,Dabholkar:2011ec,Banerjee:2009af}, completing the exact contribution of $\cN=2$ vector multiplets to the black hole entropy. The result is consistent with the logarithmic corrections from the on-shell computation \cite{Sen:2011ba}.

The organization of the paper are as follows. In section 2, we describe $\mathcal N=2$ vector multiplets on Euclidean background by taking the Euclidean continuation starting from Minkowskian supergravity. We present two possible integration contour using further analytic continuation for well defined Euclidean path integral.  We then take the AdS$_2\times$ S$^2$ background and describe the supersymmetry algebra with a choice of localization supercharge.  In section 3, we present the localization Lagrangian and the solution of localization equations. In section 4, we compute the one-loop determinant about the localization background by computing the index using Atiyah-Bott fixed point formula. In this section, we assume the trivial functional integration measure and obtain our result in terms of unphysical metric. In the next section, we propose the form of the scale invariant path integral measure and reconsider the calculation of the one-loop determinant, and therefore our main result is expressed in terms of physical variables.  We end our paper by pointing out issues and open problems in the discussion section.
\\\\
{\bf
Note added}: While this paper was being prepared for publication, we
received communication from S. Murthy and V. Reys  of a paper
which contains overlapping results \cite{Murthy:2015yfa}.

\section{$\mathcal N=2$  vector multiplets}
\subsection{Euclidean continuation}
In order to get off-shell $\cN=2$ vector multiplets in Euclidean background, we start from $\cN=2$ conformal supergravity coupled to $n_v +1$ vector multiplets by setting the Weyl multiplet as a background. Here we also translate the Lorenzian signature to Euclidean signature. For the details of the conformal supergravties, convention of gamma matrices, spinors and relation to those of Euclidean signature, we refer to the appendix \ref{Gamma} and \ref{offsugra}. 

Let us see how the fermionic fields are translated to those in Euclidean signature. 
Since the 4 dimensional Euclidean space does not allow the Majorana spinor representation,  it is useful to redefine fields in such a way that they satisfy the symplectic Majorana condition.  For the chiral and anti-chiral projection of the gaugino, poincare supersymmetry parameter and conformal supersymmetry parameter,  we use following redefinition,
\be\ba{ll}
\Omega_{i}\rightarrow \varepsilon_{ij}\lambda^{j}~~~~&~~~~\Omega^{i}\rightarrow -i \bar{\lambda}^{i}
\\
\epsilon^{i}\rightarrow \xi^{i}\,,~~~~&~~~~\epsilon_{i}\rightarrow i\varepsilon_{ij}\bar{\xi}^{j}\,,\\
\eta_{i}\rightarrow  i\varepsilon_{ij}\eta^{j}~~~~&~~~~\eta^{i}\rightarrow \bar{\eta}^{i}\,,\\
\ea\ee
where although we keep using four component notation, we use unbarred and barred notation to denote chiral and anti-chiral projected spinors. The symplectic Majorana condition in Minkowski space is 
\be
(\Psi^{i})^{\dagger}\gamma_0=-i\epsilon_{ij}(\bar{\Psi}^{j})^TC_{-}\,,~~~~~~(\bar{\Psi}^{i })^{\dagger}\gamma_0=-i\epsilon_{ij}(\Psi^{j})^T C_{-}\,,
\label{SMWM}\ee
where $C_- $ is the charge conjugation matrix. They satisfy
\be\ba{ccc}
\gamma_{a}^{T}= C_{-}\gamma_{a}C_{-}^{-1}\,,~&~C_{-}^{T}=-C_{-}\,,~&~C_{-}^{\dagger}=C_{-}^{-1}\,.
\ea\ee
Note that the chiral and anti-chiral projection is not compatible with the (symplectic) Majorana condition, so the condition (\ref{SMWM}) relates the chiral spinors and anti-chiral spinors. 
 
After hiding $\dagger$ operation on all spinors in the theory using the  symplectic Majorana condition (\ref{SMWM}),  the action and the supersymmetry transformation rule do not distinguish whether they are of Minkowkian or Euclidean theory. So, 
 we are free to go to the Euclidean theory by taking analytic continuation
\be
t= -i \theta\,.
\ee
However, we note that the property of the fermions under the complex conjugation is different. 
In the Euclidean $4$-dimensional space, we treat the chiral and anti-chiral spinors as independent fields, as they are no longer related by the complex conjugate. Instead, we can impose the following reality condition, i.e. symplectic Majorana condition, for each chiral and anti-chiral spinors,
\be
(\Psi^{i})^{\dagger}=\Psi_{i}\,,~~~~~~(\bar{\Psi}^{i })^{\dagger}=\bar{\Psi}_{i}\,,
\label{SMWEE}\ee
where the  spinors with lower $SU(2)$ index is defined as
\be
\Psi_{i}\equiv-i\epsilon_{ij}(\Psi^{j})^T C_{-}\,,~~~~~~~\bar{\Psi}_{i}\equiv -i\epsilon_{ij}(\bar{\Psi}^{j})^T C_{-}\,.
\ee
However, 
while we will choose the Killing spinors for the supersymmetric localization to satisfy this reality condition,  spinor fields may not strictly follow this condition because we will further impose analytic continuation in such a way that the path integration is well defined. 

The killing spinor equations are obtained from the variation of the gravitino,
\beqa\label{MainKilling}
&&2D_{\mu}\xi^{i}-\frac{1}{16}\gamma_{ab}T^{ab}\gamma_{\mu}\bar{\xi}^{i}-\gamma_{\mu}\bar{\eta}^{i}=0\nonumber\,,\\
&&2D_{\mu}\bar{\xi}^{i}-\frac{1}{16}\gamma_{ab}\bar{T}^{ab}\gamma_{\mu}\xi^{i}-\gamma_{\mu}\eta^{i}=0\,.
\eeqa
Here $T_{ab}$ and $\bar{T}_{ab}$ are self-dual and anti-self-dual auxiliary tensor in Weyl multiplet\footnote{For convenience, we redefine the tensor $T^{\pm}_{ab}$ in Lorenzian theory as $T^{-}_{ab}= iT_{ab}$ and $T^{+}_{ab}=i\bar{T}_{ab}$. }. And the covariant derivative includes gauge fields of both $SU(2)_R$ and $U(1)_R$. These equations determine $\eta^i$ in terms of killing spinors,
\be
\bar{\eta}^{i}=\half \slashed{D}\xi^{i}\,,~~~~~\eta^{i}=\half\slashed{D}\bar{\xi}^{i}\,.
\ee
 We also read off the auxiliary equations from variation of the auxiliary fermionic fields, $\chi^{i}$ and $\phi^{i}_{\mu}$, in the Weyl multiplet,
\beqa\label{AuxiliaryKilling}
&&-\frac{1}{24}\gamma_{ab}\slashed{D}T^{ab}\bar{\xi}^{i}+ D\xi^{i}+\frac{1}{24}i T_{ab}\gamma^{ab}\eta^{i}=0\nn\\
&&-\frac{1}{24}\gamma_{ab}\slashed{D}\bar{T}^{ab}\xi^{i}+ D\bar{\xi}^{i}+\frac{1}{24}i \bar{T}_{ab}\gamma^{ab}\bar{\eta}^{i}=0\nn\\
&&2 f_{\mu}^{a}\gamma_a \xi^{i}  +\frac{1}{16}\slashed{D}T_{ab}\gamma^{ab}\gamma_{\mu}\bar{\xi}^{i}-2D_{\mu}\bar{\eta}^{i}=0\\
&&2 f_{\mu}^{a}\gamma_a \bar{\xi}^{i}  +\frac{1}{16}\slashed{D}\bar{T}_{ab}\gamma^{ab}\gamma_{\mu}\xi^{i}-2D_{\mu}\eta^{i}=0\nn.
\eeqa
\subsection{Vector multiplets in Euclidean theory and analytic continuation}
In this section, we present the vector multiplets in Euclidean theory that is compatible with $\cN=2$ supersymmetry, and then take the analytic continuation for the contour of the path integration. 

$\mathcal N=2$ vector multiplet consist of scalars $X$ and $\bar X$, one vector field $W_\mu$, $SU(2)_R$ triplet auxiliary field $Y_{ij}$ and $SU(2)_R$ doublet fermion $\lambda^i$. For our purpose of extremal black hole, we will only consider abelian vector multiplets. The supersymmetry transformations of the vector multiplet fields are given by
\beqa\label{susyVec}\
&&\cQ X^{I}=- i\xi_{i}\,\lambda^{Ii}\nn\,,\\
&&\cQ \bar{X}^{I}=- i\bar{\xi}_{i}\,\bar{\lambda}^{Ii}\nn\,,\\
&&\cQ\lambda^{I i}= 2 i \gamma^{a}D_{a}X^{I}\bar{\xi}^{i}+\half \cF^{I}_{ab}\gamma^{ab}\xi^{i}+Y_{kj}^{I}\varepsilon^{ji}\xi^{k}+2iX\eta^{i}\,,\\
&&\cQ\bar{\lambda}^{I i}= 2 i \gamma^{a}D_{a}\bar{X}^{I}\xi^{i}+\half \cF^{I}_{ab}\gamma^{ab}\bar{\xi}^{i}+Y^{I}_{kj }\varepsilon^{ji}\bar{\xi}^{k}+2i\bar{X}\bar{\eta}^{i}\nn\,,\\
&&\cQ W_{\mu}^{I}= -\bar{\xi}_{i}\gamma_{\mu}\lambda^{iI}
-\xi_{i}\gamma_{\mu}\bar{\lambda}^{iI}\nn\,,\\
&&\cQ Y_{ij}^{I}=2\bar{\xi}_{(i}\slashed{D}\lambda^{kI}\varepsilon_{j)k}+2\xi_{(i}\slashed{D}\bar{\lambda}^{kI}\varepsilon_{j)k}\nn
\,.
\label{susy}\eeqa
where the covariant derivatives are
\beqa
&&D_{\mu}X^I= \partial_{\mu}X^I-A_{\mu}X^{I}\nn\,,\\
&&D_{\mu}\bar{X}^I= \partial_{\mu}\bar{X}^I+A_{\mu}\bar{X}^{I}\nn\,,\\
&&D_{\mu}\lambda^{iI}=(\partial_{\mu}+\quarter \omega_{\mu ab}\gamma^{ab}-\half  A_{\mu})\lambda^{iI}+\half \cV_{\mu }{}^{i}{}_{j}\lambda^{jI}\,,\\
&&D_{\mu}\bar{\lambda}^{iI}=(\partial_{\mu}+\quarter \omega_{\mu ab}\gamma^{ab}+\half  A_{\mu})\bar{\lambda}^{iI}+\half \cV_{\mu}{}^{i}{}_{j}\bar{\lambda}^{jI}\nn\,,
\eeqa
and $\cF_{\mu\nu}$ is defined as
\be\ba{l}
\cF_{\mu\nu}^{I}=F_{\mu\nu}^{I}-\quarter i\bar{X}^{I}T_{\mu\nu}
-\quarter i X^{I}\bar{T}_{\mu\nu } \,.
\label{fieldstrength}\ea\ee

The square of the supersymmetry transformations are give by
\beqa\label{susysquare}
&&\cQ^2 X^{I}=\upsilon^{\mu}D_{\mu}X^I+ \l(w + \Theta\r)X^I\nn \,,\\
&&\cQ^2 \bar{X}^{I}=\upsilon^{\mu}D_{\mu}\bar{X}^I+ \l(w - \Theta\r)\bar{X}^I \nn\,,\\
&&\textstyle{\cQ^2\lambda^{I i}=\upsilon^{\mu}D_{\mu}\lambda^{Ii}-\quarter L_{ab} \gamma^{ab}\lambda^{Ii}+ \l(\frac{3}{2}w + i\half\Theta\r)\lambda^{Ii}  +\Theta^{i}{}_{j}\lambda^{Ij}\,,}\\
&&\textstyle{\cQ^2\bar{\lambda}^{I i}=\upsilon^{\mu}D_{\mu}\bar{\lambda}^{Ii}-\quarter L_{ab} \gamma^{ab}\bar{\lambda}^{Ii}+ \l(\frac{3}{2}w -i \half\Theta\r)\bar{\lambda}^{Ii}  +\Theta^{i}{}_{j}\bar{\lambda}^{Ij}\nn\,,}
\\
&&\cQ^2 W_{\mu}^{I}= \upsilon^{\nu}(F+\bar{F})^I_{\nu\mu}+\partial_{\mu}\Phi^I\nn\,,\\
&&\cQ^2 Y_{ij}^{I}=\upsilon^{\mu}D_{\mu}Y_{ij}^{I}+2w Y_{ij}^I +Y^I_{kj}\Theta^{k}{}_{i}+Y^I_{ik}\Theta^{k}{}_{j}\nn
\,,
\eeqa
where
\beqa\label{Parameter}
&&\upsilon^{\mu}=2\bar{\xi}_{i}\gamma^{\mu}\xi^{i}\,,\quad w=-\half(\eta_{i}\xi^{i}+\bar{\eta}_{i}\bar{\xi}^i)\nn\,,\\
&&\Theta=\half(-\eta_{i}\xi^{i}+\bar{\eta}_{i}\bar{\xi}^i)\,,\\
&&L^{ab}=\quarter \xi_{i}\xi^{i}\bar{T}^{ab}+\quarter\bar{\xi}_{i}\bar{\xi}^{i}T^{ab}+\half\bar{\eta}_{i}\gamma^{ab}\bar{\xi}^{i}-\half\eta_{i}\gamma^{ab}\xi^{i}\nn\,,\\
&&\Theta^{i}{}_{j}=\bar{\xi}_j\bar{\eta}^{i}-\eta_j \xi^i -\half\delta^{i}{}_{j}(\bar{\eta}_i \bar{\xi}^i -\eta_i\xi^i)\nn\,,\\
&&\Phi^I=-2i (\bar{\xi}_{i}\bar{\xi}^{i}X^I+\xi_i\xi^i \bar{X}^I)\nn\,.
\label{para0}\eeqa
The square of the supersymmetry (\ref{susysquare}) is summarized into
\be
\cQ^2 =\cL_{v}+\mbox{Scale}(w)+R_{SO(1,1)}(\hat{\Theta})+\mbox{Lorentz}(\hat{L}^{ab}) +R_{SU(2)}(\hat{\Theta}^{i}{}_{j})+ \mbox{Gauge}(\hat{\Phi}^I)\,,
\label{Qsquare}\ee
where
\be
\hat{\Theta}= -v^{\mu}A_{\mu}+\Theta\,,~~~\hat{L}^{ab}=- v^{\mu}\omega_\mu{}^{ab}+L^{ab}\,,~~~\hat{\Theta}^i{}_{j}=\half v^{\mu}\cV_{\mu}{}^{i}{}_{j}+\Theta^i{}_{j}\,,~~~\hat{\Phi}^I= -v^{\mu}W^I_{\mu}+\Phi^I\,.
\label{hatpara}\ee

Note that the reality condition  in (\ref{SMWEE}) is compatible with the supersymmetry transformation if the bosonic fields and the background Weyl multiplet satisfy
\beqa\label{ForReal}
&&(X^{I})^{*}=-X^{I}\,,~~(\bar{X}^I)^{*}=-\bar{X}^I\,,~~(Y^I_{ij})^{*}=Y^{ij I}\,,~~~(W^I_{\mu})^{*}=W^I_{\mu}\nn\\
&&(T_{ab})^{*}=T_{ab}\,, ~~(\bar{T}_{ab})^{*}=\bar{T}_{ab}\,,~~(A_{\mu})^{*}=A_{\mu}\,,~~(\cV_{\mu}{}^{i}{}_{j})^{*}\equiv \cV_{\mu i}{}^{j}=\varepsilon_{ik}\varepsilon^{jl}\cV_{\mu}{}^{ k}{}_{l}\,.
\eeqa
That is to say, the reality condition of fermions in (\ref{SMWEE}) and bosons in (\ref{ForReal}) is preserved under the supersymmetry transformation rules given in (\ref{susy}). In particular, the symmetry parameters appeared in the algebra, (\ref{para}), satisfy the following reality conditions, 
\be
(\upsilon^{\mu})^{*}=\upsilon^{\mu}\,,~~w^{*}=w\,,~~ \Theta^{*}=\Theta\,,~~(L^{ab})^{*}=L^{ab}\,,~~(\Theta^{i}{}_{j})^{*}\equiv \Theta_{i}{}^{j}=\varepsilon_{ik}\varepsilon^{jl}\Theta^{k}{}_{l}\,,~~(\Phi^{I})^{*}=\Phi^I\,.
\ee
Therefore, the reality condition of all the fields is preserved. 
Here, the fact that parameter $\Theta$ is real reflects that the abelian factor of the R-symmetry group for the Euclidean space is $SO(1,1)_R$, whereas the $U(1)_R$ is for the Minkowskian space.

However,  we may have to take further analytic continuation. As the Eulclidean Lagrangian is of the form $\cL^{E}\sim \partial_{\mu}\bar{X}\partial^{\mu}X -Y_{ij}Y^{ij}$ which is not positive definite, the path integration is ill-defined. One natural way is to take  the path integral contour to follow \cite{Pestun:2007rz,Hama:2012bg}
\be
(X^I)^{*}=\bar{X}^I\,,~~~ (Y^I_{ij})^{*}=-Y^{ijI}\,,
\label{Contour}\ee
that make the Euclidean action positive definite. 
In this analytic continuation, the abelian R-symmetry is $U(1)_R$ as of the Minkowskian theory\footnote{ The abelian R-symmetry gauge fields should satisfy  $(A_{\mu})^{*}=-A_{\mu}$.}.
Another way is to use the localization action $-t\cQ V$ as a regulator by taking $t\rightarrow \infty$. Here we can allow the physical action not being positive definite, but still positivity on the localization saddle point is required. This way is motivated by the choice of the contour in \cite{Dabholkar:2010uh,Gupta:2012cy},
\be
(X^I)^*=X^I\,,~(\bar{X}^I)^*=\bar{X}^I\,,~Y^I_{11}= -iK^I_2 e^{i\alpha}\,,~Y^I_{22}=iK_1e^{i\beta}\,,~Y^I_{12}=Y^I_{21}=K^I_{3}\,,
\label{Contour2}\ee
where $K^I_{1,2,3}$ are real and $\alpha$ and $\beta$ are appropriately chosen coordinate dependent phase. The localization saddle point was obtained, and it turns out  the physical action  on the localization manifold is positive. 
Both of the reality conditions are not compatible with the supersymmetry transformation. The square of $\cQ$ gives rise to a gauge transformation with the parameter $\Phi$ as in the algebra (\ref{para0}) and it is not real value for both of (\ref{Contour}) and (\ref{Contour2}). Nevertheless, the argument of localization still holds because the action is invariant under supersymmetry transformations \cite{Pestun:2007rz}. 

Throughout this paper, we will be considering the first choice of the reality condition, (\ref{Contour}).  Nevertheless,  we will argue that two choices are consistent, giving same result.

\subsection{Supersymmetry on AdS$_2 \times$S$^2$}
In 4-dimensions, a supersymmetric extremal black hole has near horizon geometry of the form AdS$_2\times$S$^2$. Also all other field configurations at the near horizon are consistent with the isometry of the AdS$_2\times$S$^2$.  In the quantum entropy function, this background serves as the boundary condition for fields in the path integral.
In the Lorentzian signature, AdS$_2 \times$S$^2$ geometry implies the following ansatz,
\beqa\label{background}
&&e_{t}{}^{1}=\R\sqrt{ (r^2-1)}\,,~~e_{r}{}^{2}=\R\sqrt{1/(r^2-1)}\,,~~e_{\phi}{}^{3}=\R\sin\psi\,,~~e_{\psi}{}^{4}=\R\,,\nn\\
&&D=0\,,~~F_{rt}^{I}=e^{I}_{*}\,,~~F^{I}_{\psi\phi}=-p^{I}\sin\psi\,,~~X^{I}=X^{I}_{*}\,,~~
Y^{I}_{ij}=0\,,~~T^{-}_{rt}=\R^2\omega\,.
\eeqa
And by the attractor equations, the constant $\R$ and $ X^I_{*}$ are fixed in terms of the electric field and magnetic charges, $e^{I}_{*}$ and $ p^{I}_{*}$, and the complex constant $\omega$,
\beqa
&&\R^2=\frac{16}{\bar{\omega}\omega}\,,\nn\\
&&4(\bar{\omega}^{-1}\bar{X}^{I}_{*}+\omega^{-1}X^{I}_{*})=e^{I}_{*}\,,\\
&&4i(\bar{\omega}^{-1}\bar{X}^{I}_{*}-\omega^{-1}X^{I}_{*})=p^{I}\,\nn.
\eeqa
Solving the above equations fixes the value of the scalar field $X^I_*$ in terms of electric field and magnetic charge,
\be
{X}^I_{*}=\frac{\omega}{8}(e^I_{*}+ip^I)\,,~~~~~\bar{X}^I_{*}=\frac{\bar{\omega}}{8}(e^I_{*}-ip^I)\,.
\ee
Using the global ${U}(1)_{R}$ rotation from the superconformal Weyl multiplet, we will set  $\omega=\bar{\omega}=4/\R$. Thus with this choice of $\omega$ and $\bar{\omega}$, the $U(1)_R$ symmetry is explicitly broken. \\~\\
In the Euclidean AdS$_2\times$S$^2$ case, the near horizon field configurations take following form
\beqa\label{backgroundE}
&&e_{\theta}{}^{1}=\R\sinh{\eta}\,,~~e_{\eta}{}^{2}=\R\,,~~e_{\phi}{}^{3}=\R\sin\psi\,,~~e_{\psi}{}^{4}=\R\,\nn,\\
&&F_{ \theta \eta}^{I}=i \sinh(\eta)e^{I}_{*}\,,~~~F^{I}_{\phi\psi}=p^{I}\sin\psi\,,~~~X^{I}=X^{I}_{*}\,,~~~Y^{I}_{ij}=0\,,\\
&&D=0\,,~~~T_{\eta\theta}=-4\sinh(\eta)\R\,,~~~\bar{T}_{\eta\theta}=-4\sinh(\eta)\R\,.\nn
\eeqa
In the above we have used the $r=\cosh\eta$.
With the above vielbein, the non vanishing component of the spin connections are
\be
\omega_{\theta}^{12}=\cosh(\eta)\,,~~~~~~~~\omega_{\phi}^{34}=\cos(\psi)\,.
\ee

The background value of (\ref{backgroundE}) implies that the auxiliary Killing spinor equations (\ref{AuxiliaryKilling}) become
\be\ba{l}
\frac{1}{24}i T_{ab}\gamma^{ab}\eta^{i}=\frac{1}{24}i \bar{T}_{ab}\gamma^{ab}\bar{\eta}^{i}=0\,,\\
-2D_{\mu}\bar{\eta}^{i}= -2D_{\mu}\eta^{i}=0\,,\\
\ea\ee
which imply that $\eta^i=0$ and $\bar{\eta}^i=0$. Then, the main Killing spinor equations (\ref{MainKilling}) become
\be\ba{l}
0=2D_{\mu}\xi^{i}-\frac{1}{16}\gamma_{ab}T^{ab}\gamma_{\mu}\bar{\xi}^{i}\,,\\
0=2D_{\mu}\bar{\xi}^{i}-\frac{1}{16}\gamma_{ab}\bar{T}^{ab}\gamma_{\mu}\xi^{i}\,.\\
\ea\ee
It is solved in \cite{Lu:1998nu} and there are $8$ Killing spinors. For the purpose of the supersymmetric localization,  we will choose the following two Killing spinors among them. 
In terms of Dirac spinor notation,
\be
\zeta^{i}:= \xi^{i}+\bar{\xi}^{i}\,,
\ee
and in the following gamma matrix representation,
\be
\gamma^{1}=\sigma_1 \otimes 1\,,~~~\gamma^{2}=\sigma_2 \otimes 1\,,~~~\gamma^{3}=\sigma_3 \otimes \sigma_1\,,~~~\gamma^{4}=\sigma_3\otimes\sigma_2\,,~~~\gamma_5=-\gamma_{1234}=\sigma_3\otimes\sigma_3\,,
\ee
our choice of the Killing spinors are
\be
\zeta^{1}=\sqrt{2} e^{i(\theta+\phi)/2}\left( \ba{c} \sinh\frac{\eta}{2} \sin\frac{\psi}{2} \\ \cosh\frac{\eta}{2} \sin\frac{\psi}{2}\\ \sinh\frac{\eta}{2}\cos\frac{\psi}{2}\\ \cosh\frac{\eta}{2}\cos\frac{\psi}{2}\ea\right)\,,~~~~~~
\zeta^{2}=\sqrt{2}  e^{-i(\theta+\phi)/2}\left( \ba{c} \cosh\frac{\eta}{2} \cos\frac{\psi}{2} \\ \sinh\frac{\eta}{2} \cos\frac{\psi}{2}\\ -\cosh\frac{\eta}{2}\sin\frac{\psi}{2}\\- \sinh\frac{\eta}{2}\sin\frac{\psi}{2}\ea\right)\,.
\ee
These Killing spinors satisfy the  the symplectic Majorana condition
\be
(\zeta^{1})^{*}=-i \varepsilon_{12}(\sigma_{1}\otimes \sigma_{2})\zeta^{2}\,,~~~~\varepsilon_{12}=1\,.
\ee

Our choice of charges generates the killing vector field of the compact isometry transformation as
\be
2(\bar{\xi}_{i}\gamma^{\mu}\xi^{i})\partial_{\mu}=4\frac{1}{\R}(\partial_{\theta}-\partial_{\phi})=-i4(L -J)\,,
\ee 
where we denote $L$ as the rotation of the AdS$_2$ and $J$ as the rotation of the S$^2$. 
We  also note that
\be
\xi_{i}\xi^{i}= 2(\cosh \eta +\cos\psi )\,,~~~\bar{\xi}_{i}\bar{\xi}^i = 2(\cosh \eta -\cos\psi)\,.
\ee
Then the symmetry parameters (\ref{Parameter}) are given by
\beqa
&&v^{\mu}\partial_{\mu}=2(\bar{\xi}_{i}\gamma^{\mu}\xi^{i})\partial_{\mu}=4\frac{1}{\R}(\partial_{\theta}-\partial_{\phi})\,,\nn\\
&&L^{12}= \half\cosh(\eta)(\omega +\bar{\omega})+\half \cos{\psi}(\bar{\omega}-\omega)=4\frac{1}{\R}\cosh(\eta)\,,\nn\\
&&L^{34}=\half\cosh(\eta)(\omega-\bar{\omega})-\half\cos(\psi)(\omega+\bar{\omega})=-4\frac{1}{\R}\cos(\psi)\,,\\
&&\Phi^I=-4 i \cosh(\eta) X^I_1-4 \cos (\psi) X^I_2\,,\nn\\
&&w=\Theta=\Theta^{i}{}_{j}=0\,,
\eeqa
where
\be
X^I_1\equiv X^I+ \bar{X}^I\,,~~~ X^I_2 \equiv -i (X^I-\bar{X}^I)\,.
\ee
Therefore, the parameters (\ref{hatpara}) in the supersymmetry algebra (\ref{Qsquare}) are
\beqa\label{para}
&&\hat{\Theta}= -v^{\mu}A_{\mu}=0\,,~~~\hat{L}^{12}= 0\,,~~~\hat{L}^{34}= 0\,,\nn\\
&&\hat{\Theta}^i{}_{j}=\half v^{\mu}\cV_{\mu}{}^{i}{}_{j}=0\,,~~~\hat{\Phi}^I= -v^{\mu}W^I_{\mu}-4 i \cosh(\eta) X^I_1-4\cos (\psi) X^I_2\,.
\eeqa
Note here that $v^{\mu}A_{\mu}=0$ and $ v^{\mu}\cV_{\mu}{}^{i}{}_{j}=0$ as the background value of them are zero. It  still holds when we consider the Weyl multiplet as localization saddle point \cite{Gupta:2012cy}.
\section{Localization}
We deform the physical action by adding the following localization Lagrangian,
\be
\cL^{\cQ}=\cQ V\,,~~~~~~V= ({\xi_{j}\xi^{j}+\bar{\xi}_{j}\bar{\xi}^{j}})^{-1} \sum_{I=0}^{n_{V}}(\cQ\lambda^{iI})^{\dagger}\lambda^{iI}+(\cQ \bar{\lambda}^{iI})^{\dagger}\bar\lambda^{iI}\,.
\label{QV}\ee
Here, we take overall normalization factor $({\xi_{j}\xi^{j}+\bar{\xi}_{j}\bar{\xi}^{j}})^{-1}$ such that we will get standard kinetic terms for scalars and fermions. 
 Note that the localization Lagrangian is by construction positive definite as it involves the dagger operation.
The dagger operation should be taken  carefully because it relies on which contour of integration that we choose. 
For the positive definiteness of the Euclidean action, we gave up the the reality condition (\ref{SMWEE}) for fermions and performed further analytic continuation, following the contours defined in (\ref{Contour}).
\subsection{Localization saddle points}
To look at the localization saddle point, let us consider the bosonic part of the localization Lagrangian. After some algebra, one can rewrite the bosonic part of the localization Lagrangian (\ref{QV}) as follows\footnote{Here we set the $U(1)_R$ gauge field, $A_{\mu}$, to be zero as the localization saddle point in Weyl multiplet \cite{Gupta:2012cy}.},
\be\ba{ll}
({\xi_{j}\xi^{j}+\bar{\xi}_{j}\bar{\xi}^{j}}) \cL^{\cQ}_{b}=&\quarter\left(\frac{1}{\xi_{i}\xi^{i}}+\frac{1}{\bar{\xi}_{i}\bar{\xi}^{i}}\right)\left[ (v^{\mu}\partial_{\mu} X_1^{I})^2+(v^{\mu}\partial_{\mu} X_2^{I})^2\right]\\
&+\xi_{i}\xi^{i}\left| F^{I+}_{ab}-\frac{1}{8}X_2^{I} T_{ab}+\frac{1}{\xi_{j}\xi^{j}}v_{[a}\partial_{b]+}X_2^{I}\right|^2\\&+\bar{\xi}_{i}\bar{\xi}^{i}\left| {F}_{ab}^{I-}+\frac{1}{8}X_2^{I} \bar{T}_{ab}-\frac{1}{\bar{\xi}_{j}\bar{\xi}^{j}}v_{[a}\partial_{b]-}X_2^{I}\right|^2\\
&+\half \xi_l\xi^l \left| iY^{I}_{ik}\varepsilon^{kj} +\frac{1}{8 \xi_k \xi^k}X_{1}^{I}T_{ab}\xi_{i}\gamma^{ab}\xi^{j}+\frac{1}{(\xi_{k}\xi^{k})^2}v_{[a}\partial_{b]+}X^{I}_{1}\xi_{i}\gamma^{ab}\xi^{j}\right|^{2}\\
&+\half \bar{\xi}_l\bar{\xi}^l \left| i Y^{I}_{ik}\varepsilon^{kj} +\frac{1}{8 \bar{\xi}_k \bar{\xi}^k}X_{1}^{I}\bar{T}_{ab}\bar{\xi}_{i}\gamma^{ab}\bar{\xi}^{j}+\frac{1}{(\bar{\xi}_{k}\bar{\xi}^{k})^2}v_{[a}\partial_{b]-}X_{1}^{I}\bar{\xi}_{i}\gamma^{ab}\bar{\xi}^{j}\right|^{2}\,,
\label{bosonicL}\ea\ee
where the symbol $\pm$ indicates the self-dual or anti-self-dual parts such that,
\be
F^{I\pm}_{ab}=\half (F^I_{ab}\pm\half \varepsilon_{abcd}F^{Icd})\,,\quad v_{[a}\partial_{b]\pm}:= \half (v_{[a}\partial_{b]}\pm\half \varepsilon_{abcd}v^{[c}\partial^{d]})\,.
\ee
$\bullet$ {\bf The localization solution }: The condition $0=\cL^{\cQ}_{b}$ gives $6$ localization saddle point equations. Classical background is the trivial solution. On top of this we find the other off-shell solutions. From the first and the last two lines in (\ref{bosonicL}), one finds the unique solution for $X^I_1$ and $Y^I_{ij}$ \cite{Dabholkar:2010uh,Gupta:2012cy},
\be
X_{1}^{I}=\frac{C_1^{I}}{\R\cosh \eta}\,,~~~~~~Y^{I}_{12}=-\frac{C_1^{I}}{ \R^2 \cosh^{2} \eta}\,,~~~~~Y^{I}_{11}=Y^{I}_{22}=0\,.
\label{solutionX1}\ee
Similarly first three lines of (\ref{bosonicL}) provides equations for $F^I_{ab}$ and $X^I_2$. We find a nontrivial smooth solution which is given by
\beqa\label{solutionX2}
&&\displaystyle{
X_{2}^{I}=\frac{C_2^{I}\cos\psi}{\R\cosh\eta}\,,~~~F^{I}_{12}=-\frac{C_2^{I}}{\R^2 \cosh^2\eta}\,,~~~F^{I}_{23}=\frac{C_2^{I}\sinh\eta\sin\psi}{ \R^2\cosh^2\eta}\,,~~~F^{I}_{34}=\frac{2C_2^{I}\cos\psi}{\R^2\cosh\eta}\,,
}\nn\\
&&\,\,\,\,\, \qquad \displaystyle{F_{13}=F_{24}=F_{14}=0\,.}
\label{X2zero}
\eeqa
However, we cannot prove that this is the unique smooth solution but we will provide evidence in support of it in the appendix \ref{X2sol}.  One important feature of the above solution is that although there is a non trivial field strength along S$^2$, the total flux however is zero. Thus the magnetic charge for this off-shell solution is same as the attractor value.\\
$\bullet${\bf  At north/south pole and the origin}: At the fixed points $\eta=0$ with $\psi=0$ or $\psi=\pi$, it seems that further singular solutions can be enhanced because $\xi_{i}\xi^{i}=0$ or $\bar{\xi}_{i}\bar{\xi}^{i}=0$ at this point and the number of localization equations are reduced. However, we will argue that there are no nontrivial solutions that are localized at the fixed points. 

Consider the point $\eta=0$ and $\psi=0$. Using the fact that
\be
\bar{\xi}_{i}\bar{\xi}^{i}=0\,,~~~~~\bar{\xi}_{i}\gamma^{a}\xi^{j}=0\,,~~~~~\bar{T}_{ab}\bar{\xi}_{j}\gamma^{ab}\bar{\xi}^{i}=0\,,
\ee
and after some algebra, one finds that the localization Lagrangian reduces to
\be\ba{ll}
 \cL^{\cQ}=&\quarter \frac{1}{(\xi_{i}\xi^{i})^2}\left[ (v^{\mu}\partial_{\mu} X_1^{I})^2+(v^{\mu}\partial_{\mu} X_2^{I})^2\right]+(\partial^{\mu}X_{1}\partial_{\mu}X_{1}+\partial^{\mu}X_{2}\partial_{\mu}X_{2})\\
&+\left| F^{I+}_{ab}-\frac{1}{8}X_2^{I} T_{ab}\right|^2+\half  \left| iY^{I}_{ik}\varepsilon^{kj} +\frac{1}{8 \xi_k \xi^k}X_{1}^{I}T_{ab}\xi_{i}\gamma^{ab}\xi^{j}\right|^{2}\,.\\
\ea\ee
From the first line, we get $X^I_1$ and $X^I_2$ to be constant. Since  $e_{\theta}^{1}=e_{\phi}^{3}=0$ at the north pole, we get the anti-self dual equation from the first term of the second line,
\be
F^{I+}_{\mu\nu}= \frac{1}{8}X^I_2 \,e_{\mu}^{a}e_{\nu}^{b}T_{ab}=0\,.
\ee
Similarly, at $\eta=0$ and $\psi=\pi$, we get the self-dual equation,
\be
F^{-}_{\mu\nu}=0\,.
\ee
However, there is no $U(1)$ instantons in $4$-dimensions, so there is no localized nontrivial solutions.\\
$\bullet${\bf Boundary mode (discrete zero modes) of the gauge field}\\
Apart from the zero mode in (\ref{solutionX1}) and (\ref{solutionX2}), the AdS$_2$, which is a non-compact space, forces us to consider so called boundary modes of gauge fields \cite{Camporesi:1994ga}. 
\be
W^{l}= d \Phi^{l}\,,~~~~ \Phi^{l}=\frac{1}{\sqrt{2\pi |l|}}\left[ \frac{\sinh \eta}{1+\cosh \eta}\right]^{|l|}e^{il\theta}\,,~~~l=\pm 1,\pm2, \pm3, \cdots\,.
\label{discretemode}\ee
These modes are not actually the ``localizing saddle points'' in the sense of $\cQ$ invariant BPS states as it is obvious that $\cQ^2 \neq 0$ for those modes. 
However,  these are the zero modes making the localization action as well as the original action vanish since the filed strength is zero. Yet, these are not pure gauge modes as the parameters $\Phi^l$ are not normalizable. These modes do not vanish at the boundary of the AdS$_2$, but  they are still normalizable. Thus the integration over theses boundary modes should be taken into account for the partition function.  Although it will be infinite product of integrations, the regularized result is well understood \cite{Banerjee:2010qc}. 

\section{$1$-loop partition function}
In this section, we compute the $1$-loop partition function by computing equivariant index. For this, we introduce BRST symmetry to fix the gauge and combine it with the localization supercharge. Through out this section, we assume the ordinary path integration measure. The correct measure will be taken into account in the next section.

\subsection{BRST and combined cohomology}

$\bullet$ {\bf{Cohomological variables and supersymmetry complex}}: It is useful to present the supersymmetry in the cohomological form by changing the variables. Our fermionic variables are reorganized as
\beqa\label{Cohovariable}
&&\Psi^I\equiv \cQ X^I_2 =- \xi_{i}\lambda^{iI}+\bar{\xi}_{i}\bar{\lambda}^{iI}\,,\nn\\
&&\Psi^I_{\mu}\equiv \cQ W^I_{\mu}=-\bar{\xi}_{i}\gamma_{\mu}\lambda^{iI}-\xi_{i}\gamma_{\mu}\bar{\lambda}^{iI}\,,\\
&&\Xi^{I ij}\equiv2\xi^{(i}C_-\lambda^{j)I}+2 \bar{\xi}^{(i}C_-\bar{\lambda}^{j)I}\,.\nn
\eeqa
Then the inverse relation is
\beqa\label{inverserelation}
&&-\xi^{i}\Psi^I -\gamma^{\mu}\bar{\xi}^{i}\Psi^I_{\mu}+i\varepsilon_{jk}\xi^{k}\Xi^{I ji}=(\xi_{j}\xi^{j}+\bar{\xi}_{j}\bar{\xi}^{j})\lambda^{iI}=4\cosh(\eta)\lambda^{iI}\,, \nn\\
&&+\bar{\xi}^{i}\Psi^I -\gamma^{\mu}\xi^{i}\Psi^I_{\mu}+i\varepsilon_{jk}\bar{\xi}^{k}\Xi^{Iji}=(\xi_{j}\xi^{j}+\bar{\xi}_{j}\bar{\xi}^{j})\bar{\lambda}^{iI}=4\cosh(\eta)\bar{\lambda}^{iI} \,.
\eeqa
In terms of these variable, the supersymmetry transformations are
\beqa
&&\cQ X^I_2=\Psi^I\,,~~~~~~~\cQ\Psi^I= \cL_v X^I_2 
\,,\nn\\
&&\cQ W^I_{\mu}=\Psi^I_{\mu}\,,~~~~~~\cQ\Psi^I_{\mu}=\cL_{v}W^I_{\mu}+ \partial_{\mu}\hat{\Phi}^I
\,,~~~~~~{\cQ\hat{\Phi}^{I}=0}\,,\\
&&\cQ\Xi^{Iij}=B^{Iij}\,,~~~~\cQ B^{Iij}=\cL_v \Xi^{Iij}
\,.\nn
\eeqa
Here $\hat{\Phi}^{I}$ contains the degree of freedom $X^I_1$ as in (\ref{para}), 
and $B^{Iij}$ contains the degree of freedom  $Y^{Iij}$  as 
\be\ba{l}
B^{Iij}:= 4 \bar{\xi}^{(i}C_{-}\gamma^{\mu}\xi^{j)}\partial_{\mu}X^I_2 +i (\xi_{k}\xi^{k}+\bar{\xi}_{k}\bar{\xi}^{k})Y^{Iij}\\
~~~~~~~~~~~ +\xi^{(i} C_{-}\gamma^{ab}\xi^{j)}(F^{I+}_{ab}-\quarter i \bar{X^I}T_{ab})+\bar{\xi}^{(i} C_{-}\gamma^{ab}\bar{\xi}^{j)}(F^{I-}_{ab}-\quarter i X^I\bar{T}_{ab})\,.
\label{delPsiXi}\ea\ee
Note that all the bosonic variables are organized into $(X^I_2\,, W^I_{\mu}\,, B^{Iij}\,,{\hat{\Phi}^{I}})$ and all the fermionic variables are into $(\Psi^I\,,\Psi^I_{\mu}\,,\Xi^{Iij})$.
And the ${\cQ}^2$ acts as
\be
{\cQ}^2=\cL_{v}+\mbox{Gauge}(\hat{\Phi})\,.
\ee
In general, ${\cQ}$ could act as ${\cQ}^2=\cL_{v}+\mbox{Gauge}(\hat{\Phi})+\mbox{Lorentz}(\hat{L}_{ab})+R_{U(1)}(\hat{\Theta})+R_{SU(2)}(\hat{\Theta}^{i}{}_{j})$. However, we note from (\ref{para}) that $\hat{L}_{ab}=\hat{\Theta}=\hat{\Theta}^{i}{}_{j}=0$.  
 \\~\\
$\bullet$ {\bf BRST complex}: To treat the gauge fixing of the $U(1)^{n_{v}+1}$ Yang-Mills gauge symmetry, we introduce the ghost fields and use BRST quantization. The BRST complex is
\beqa
&&\cQ_{B}W^I_{\mu}=\partial_{\mu}c^I\,,~~~~~~\cQ_{B}c^I=0\,,\nn\\
&&\cQ_{B}\bar{c}^I=B^I\,,~~~~~~~~~\cQ_{B}B^I=0\,, \\
&&\cQ_B\lambda^{iI}=\cQ_B\bar{\lambda}^{iI}=\cQ_B X^I=\cQ_B\bar{X}^I=\cQ_B Y^I_{ij}=0\,.\nn
\eeqa
Here the $c^I\,,\bar{c}^I$ and $B^I$ are the ghost, anti-ghost and the standard Lagrange multiplier, so that the gauge fixing is performed by adding the terms, $\cL_{GF}=iB^I \nabla_{\mu}W^{I\mu }+\frac{\xi}{2}B^{I2}+\bar{c}^I \Box c^I$. 
We assign the length dimension of the BRST operator $ [\cQ_{B}]=-\half$, so the
length dimension for the ghost multiple is set by
\be
[c]=-\half\,,~~~~[\bar{c}]=-\half\,,~~~~~[B]=-1\,.
\ee
Note that the AdS$_2 \times$S$^2$ space does not have normalizable zero mode of $c^I\,,\bar{c}^I$ and $B^I$. The boundary condition of the path integral does not allow the non-normalizable modes, so we do not need special treatment for freezing out these kind of zero modes. This differs from the case of S$^4$ space.   We refer to \cite{Pestun:2007rz} as the S$^4$ example where there are constant zero modes so the additional constant fields are introduced to freeze out those modes.  

The gauge fixing Lagrangian is $\cQ_B$-exact, so
\be
\cL^{\cQ}_{GF}= \cQ_B \left[\R^{-1}\bar{c}^I i \nabla_{\mu}W^{I\mu}+\R^{-2}\frac{\xi}{2} \bar{c}^I B^I\right]\,,
\label{LQGF}\ee
where we put constants factors $\R^{-1}$ and $\R^{-2}$ to set the length dimension $-4$ for the Lagrangian.\\
~\\
$\bullet${\bf Combined complex}: Since the gauge fixing Lagrangian (\ref{LQGF}) is not $\cQ$ invariant, we need to consider new complex and modify the gauge fixing Lagrangian. Combining the BRST symmetry with supersymmetry, we make the combined complex.  For this we define the supersymmetry for the ghost
\beqa\label{susyghost}
&&\cQ c^I=-\hat{\Phi}^I\,,~~~~\cQ\hat{\Phi}^I=0\,,\nn\\
&&\cQ B^I=\cL_{v}\bar{c}^I\,,~~~~~\cQ\bar{c}^I=0\,,
\eeqa
and we introduce the combined operator $\hat{\cQ}:=\cQ+\cQ_{B}$.
 Then we get the following combined $\hat{\cQ}$-complex,
\beqa\label{complex}
&&\hat{\cQ}W^I_{\mu}=\Psi^I_{\mu}+\partial_{\mu}c^I\,,~~~\hat{\cQ}\Psi^I_{\mu}=\cL_{v}W^I_{\mu}+\partial_{\mu}\hat{\Phi}^I\,,\nn\\
&&\hat{\cQ}X^I_2=\Psi^I\,,~~~~~~~~~~~~~\hat{\cQ}\Psi^I= \cL_v X^I_2
\,,\nn\\
&&\hat{\cQ}c^I=-\hat{\Phi}^I\,,~~~~~~~~~~~~\hat{\cQ}\hat{\Phi}^I=-\cL_{v}c^I\,,\\
&&\hat{\cQ}B^I=\cL_{v}\bar{c}^I\,,~~~~~~~~~~~\hat{\cQ}\bar{c}^I=B^I\,,\nn\\
&&\hat{\cQ}\Xi^{Iij}=B^{Iij}\,,~~~~~~~~~~\hat{\cQ}B^{Iij}=\cL_v \Xi^{Iij}
\,.\nn
\eeqa
In fact supersymmetry transformation for ghost (\ref{susyghost}) was defined such that the $\hat{\cQ}^2$ acts as
\be
\hat{\cQ}^2=\cL_{v}:=H.
\ee
All the bosonic and fermionic variables are organized as 
\beqa\label{variableset}
&&\mathbb{ X}:= (X^I_2\,,W^I_{\mu})\,,~~~~~~ \hat{\cQ}\mathbb{ X}= (\hat{\cQ} X^I_2\,,\hat{\cQ} W^I_{\mu})\,,\nn\\
&&\Xi:=(\Xi^{Iij}\,, \bar{c}^I\,,c^I)\,,\quad\hat{\cQ}\Xi=(\hat{\cQ} \Xi^{Iij}\,,\hat{\cQ}\bar{c}^{I}\,,\hat{\cQ}c^{I})\,.
\eeqa

We now use $\hat{\cQ}$-exact gauge fixing term,
\be
\cL^{\hat{\cQ}}_{GF}= \hat{\cQ}(i\R^{-1}\bar{c}^I\nabla_{\mu}W^{I\mu} +\R^{-2}\frac{\xi}{2}\bar{c}^{I}B^{I})\,.
\label{LGF}\ee
This is equivalent to the  (\ref{LQGF}) as
\be
\cL^{\hat{\cQ}}_{GF}=\cL^{\cQ}_{GF} -i\R^{-1}\bar{c}^{I} \nabla_{\mu}\Psi^{I\mu} -\R^{-2}\frac{\xi}{2}\bar{c}^{I}\cL_{v}\bar{c}^{I}\,.
\ee
and the terms, $-i\bar{c}^{I} \nabla_{\mu}\Psi^{I\mu}$ and $-\frac{\xi}{2}\bar{c}^{I}\cL_{v}\bar{c}^{I}$, do not contribute to the determinant. It is because $\bar{c}$ can be connected only to $c$ but there are no vertices in those extra terms containing $c$.

Now, the physical action is $\hat{\cQ}$ invariant since it is invariant under the $\cQ$ and $\cQ_B$ symmetry.  Also the gauge fixing Lagrangian is $\hat{\cQ}$ invariant. For the supersymmetric localization, we now deform the physical action by adding the following $\hat{\cQ}$ exact terms,
\be
\hat{\cQ}\cV=\hat{\cQ}\left[ \frac{1}{({\xi_{j}\xi^{j}+\bar{\xi}_{j}\bar{\xi}^{j}}) }\left((\hat{\cQ}\lambda^{iI})^\dagger \lambda^{iI}+(\hat{\cQ}\bar{\lambda}^{iI})^\dagger \bar{\lambda}^{iI}\right)+ i\R^{-1}\bar{c}^I\nabla_{\mu}W^{I\mu} +\R^{-2}\frac{\xi}{2}\bar{c}^{I}B^{I}\right]\,.
\ee
Since $\hat{\cQ}\lambda^{iI}={\cQ}\lambda^{iI}$ and $\hat{\cQ}\bar{\lambda}^{iI}={\cQ}\bar{\lambda}^{iI}$, the localization equation obtained from (\ref{bosonicL}) will not be changed. To express it in terms of the set of the cohomological variables in (\ref{variableset}), we use the inverse relation (\ref{inverserelation}). Then we find,
\be
\hat{\cQ}\cV= \hat{\cQ}\left[\frac{1}{(4\cosh\eta)^2}\left[ (\hat{\cQ}\Psi^I)^{\dagger}\Psi^I +(\hat{\cQ} \Psi^{I\mu})^{\dagger}\Psi^I_{\mu}+ \half (\hat{\cQ} \Xi^{Iij})^{\dagger}\Xi^{Iij}\right]  + i\R^{-1}\bar{c}^I\nabla_{\mu}W^{I\mu} +\R^{-2}\frac{\xi}{2}\bar{c}^{I}B^{I}\right]\,,
\ee
where explicitly, we note that
\be
\Psi=\cQ X_2=\hat{\cQ}X_2\,,\qquad
\Psi^{I}_{\mu}=\cQ W^I_{\mu}= \hat{\cQ}W^I_{\mu}-\partial_{\mu}c\,,\qquad \hat{\cQ}\Xi^{Iij}=B^{Iij}\,,
\ee
and their conjugation are given by
\beqa\label{daggeronfields}
(\hat{\cQ}\Psi)^{\dagger}~&=&\hat{\cQ}\Psi\,,\nn\\
(\hat{\cQ} \Psi_\mu^I)^\dagger&=&\cL_v W^I_\mu + \partial_\mu [-v^{\nu}W^I_{\nu}+4i\cosh(\eta)X^I_1 -4\cos(\psi)X_2^I]\\
&=&\cL_{v}W^I_{\mu} -2\partial_{\mu}\left( v^{\nu}W^I_{\nu}+4\cos\psi X^I_2 \right) +\partial_{\mu}\left(  \hat{\cQ}c^I \right)\,,\nn
\\
(\hat{\cQ} \Xi^{Iij})^\dagger&=&-4\, \varepsilon_{ik}\varepsilon_{jl}\bar{\xi}^{(k}C_{-}\gamma^{\mu}\xi^{l)}\partial_{\mu}X^I_2 
+i(\xi_{k}\xi^{k}+\bar{\xi}_{k}\bar{\xi}^{k})Y^I_{ij}\nn\\
&&- \varepsilon_{ik}\varepsilon_{jl}\left[\xi^{(k} C_{-}\gamma^{ab}\xi^{l)}(F^{I+}_{ab}+\quarter i X^IT_{ab})+\bar{\xi}^{(k} C_{-}\gamma^{ab}\bar{\xi}^{l)}(F^{I-}_{ab}+\quarter i \bar{X}^I\bar{T}_{ab})\right]\nn\\
&=&  \varepsilon_{ik}\varepsilon_{jl} \left[ \hat{\cQ}\Xi^{kl} - 8 \bar{\xi}^{(k}C_{-}\gamma^{a}\xi^{l)}\partial_{a}X^I_2 
 \right.\nn\\
&&\left.-\xi^{(k}C_{-} \gamma^{ab}\xi^{l)}(2F^{I+}_{ab}-\quarter  X^I_{2}T_{ab})-\bar{\xi}^{(k} C_{-}\gamma^{ab}\bar{\xi}^{l)}(2F^{I-}_{ab}+\quarter  X^I_{2}\bar{T}_{ab})\right]\,.\nn
\eeqa
\subsection{Index and 1-loop determinant}\label{1-loop}
To evaluate the $1$-loop determinant, We formally write the quadratic terms of the localization Lagrangian, in terms of the new variable set (\ref{variableset}), as
\be
\hat{\cQ}\cV = \hat{\cQ}\left[(\hat{\cQ} \mathbb{X}'\,, \Xi')\begin{pmatrix}D_{00}&D_{01}\\D_{10}&D_{11}\end{pmatrix}\begin{pmatrix}\mathbb{X}'\\\hat{\cQ}\Xi' \end{pmatrix}\right]\,.
\label{formalQV}\ee
Here, we denoted  $(\mathbb{X}'\,,\hat{\cQ}\mathbb{X}'\,, \Xi' \,, \hat{\cQ} \Xi')$ to  exclude the zero modes, yet $\mathbb{X}'$ is to include the boundary modes of the gauge field (\ref{discretemode})\footnote{ Although the boundary gauge modes are zero modes, the corresponding fermion modes  in $\hat{\cQ}\mathbb{X}'$ are not zero modes. It is known that there is no such infinite set of fermionic zero modes \cite{Banerjee:2011jp}. }.

Among the fluctuation modes of $\mathbb{X}'$ and $\Xi'$, some of them can be annihilated by $\hat{\cQ}^2=H$. Let us classify the set of the path integration variable into two parts,
\be\ba{ll}
\mathbb{X}''=\{\mathbb{X}'| H \mathbb{X}'\neq 0 \}\,,~~~~~&\Xi''=\{\Xi' | H \Xi'\neq 0 \}\,,\\
\mathbb{X}'^{0}=\{\mathbb{X}'| H \mathbb{X}'=0 \}\,,&\Xi'^0=\{\Xi' | H \Xi'= 0 \}\,.\\
\ea\ee
Then, since $H$ commutes with $\hat{\cQ}$ and $D_{ij}$, the terms in the localization Lagrangian can be separated as
\be\ba{ll}
\hat{\cQ}\cV
&= (\mathbb{X}''\,, \hat{\cQ}\Xi'')K''_b\begin{pmatrix}\mathbb{X}''\\
\hat{\cQ}\Xi'' \end{pmatrix}
+ (\mathbb{X}'^{0}\,, \hat{\cQ}\Xi'^{0})K'^{0}_b\begin{pmatrix}\mathbb{X}'^{0}\\
\hat{\cQ}\Xi'^{0} \end{pmatrix}\\
&~+ 
(\hat{\cQ} \mathbb{X}''\,, \Xi'')K''_f \begin{pmatrix}\hat{\cQ}\mathbb{X}''\\\Xi'' \end{pmatrix}+ 
(\hat{\cQ} \mathbb{X}'^{0}\,, \Xi'^{0})K'^{0}_f \begin{pmatrix}\hat{\cQ}\mathbb{X}'^{0}\\\Xi'^{0} \end{pmatrix}\,,
\ea\ee
where the kinetic operators of bosons and fermions, $K_b$ and $K_f$, are divided as
\be\ba{ll}
K''_{b}= \begin{pmatrix}-H&0\\0&1\end{pmatrix}\!\!\begin{pmatrix}D_{00}&D_{01}\\D_{10}&D_{11}\end{pmatrix}\!+\!\begin{pmatrix}D_{00}^{T}&D_{10}^{T}\\D_{01}^{T}&D_{11}^{T}\end{pmatrix}\!\!\begin{pmatrix}H&0\\0&1\end{pmatrix}\,,&K'^{0}_b= \begin{pmatrix}0 &D_{10}^{T}\\D_{10}&D_{11}\!+\!D_{11}^{T}\end{pmatrix}\,,\\
K''_{f}=\begin{pmatrix}1&0\\0&-H\end{pmatrix}\!\!\begin{pmatrix}D_{00}^{T}&D_{10}^{T}\\D_{01}^{T}&D_{11}^{T}\end{pmatrix}\! -\!\begin{pmatrix}D_{00}&D_{01}\\D_{10}&D_{11}\end{pmatrix}\!\!\begin{pmatrix}1&0\\0&H\end{pmatrix}\,,&K'^{0}_f= \begin{pmatrix}D_{00}^{T}\!-\!D_{00} &D_{10}^{T}\\-D_{10}&0\end{pmatrix}\,.
\ea
\ee
Note that the determinant of $K'^{0}_b$ and $K'^{0}_{f}$ cancels with each other since $D_{10}$ is non degenerate for the corresponding modes. Also note that
\be
\begin{pmatrix}1&0\\0&-H\end{pmatrix} K''_{b}= K''_{f}\begin{pmatrix}H&0\\0&1\end{pmatrix}\,.
\ee
So the $1$-loop determinant is given, up to a sign, by
\be
Z_{1-loop}=\left(\frac{\det{}' K_{f}}{\det{}' K_{b}}\right)^{1/2}=\left(\frac{\det{}'' K''_{f}}{\det{}'' K''_{b}}\right)^{1/2}= \left(\frac{\det_{\hat{\cQ}\Xi''}H}{\det_{\hat{\cQ}\mathbb{X''}}H}\right)^{1/2}= \left(\frac{\det_{\Xi''}H}{\det_{\mathbb{X''}}H}\right)^{1/2}\,,
\label{detH}\ee
where the last equality is due to that $\hat{\cQ}$ commutes with $H$. 

The $1$-loop determinant expressed in (\ref{detH}) is encoded in the following quantity, 
\be\ba{l}
\Tr_{\mathbb{X}''}e^{tH}-\Tr_{\Xi''}e^{tH}\,,
\ea\label{Trace1}\ee
which will be expressed as a formal Laurent series in $U(1)$ representation, i.e. $e^{it/\ell}$. 
Calculating this, we can read off the eigenvalues and the degeneracies and then obtain the ratio of the determinant as 
\be
\sum_{n} \omega_n e^{it \varepsilon_{n}} \longrightarrow  \frac{\det_{\mathbb{X''}}H}{\det_{\Xi''}H} =\prod_n (\varepsilon_n)^{\omega_n}\,.
\ee

To compute the (\ref{Trace1}), it is convenient to express the trace as the summation over the complete set of basis.  Firstly, we freely add the trace over  $\mathbb{X}'^{0}$ and $\Xi'^{0}$.   Their contributions cancel each other since operator $D_{10}$ maps the fields $\mathbb{X}$ to the dual of the fields $\Xi$ and it is non-degenerate for those mode. Secondly, we add and subtract possible zero mode contributions in $\mathbf{X}$ and $\Xi$. Then  the (\ref{Trace}) becomes
\be\ba{l}
\Tr_{\mathbb{X}}e^{tH}-\Tr_{\Xi}e^{tH} -N^0_{\mathbb{X}} + N^0_{\Xi}\,.
\ea\label{Trace}\ee
Since (\ref{Trace1}) does not give $t$-independent constant, (\ref{Trace}) does not either. It will turn out that the number $N^0_{\mathbb{X}} - N^0_{\Xi}$ should vanish as the first two term will not produce $t$-independent constant later in (\ref{indD10}).  As we have a single zero mode (\ref{X2zero}) in $\mathbb{X}$, i.e. $N^0_{\mathbb{X}}=1$,  we have a single fermion zero mode in $\Xi$, i.e. $N^0_{\Xi}=1$ \footnote{In case of another analytic continuation (\ref{Contour2}), we would get $N^0_{\mathbb{X}}=N^0_{\Xi}=0$.} \footnote{The fermion mode should appear in pair, and the other fermion zero mode is in $\hat{\cQ}\mathbb{X}$. One can easily see that the explicit solution is $\Psi =\vartheta \cos\psi/\cosh\eta\,,\Psi_{\theta}=-\vartheta/\cosh\eta\,,\Psi_{\phi}=-\vartheta\sin^2\psi/\cosh\eta\,,\Psi_{\eta}=\Psi_{\psi}=0$ with the grassman parameter $\vartheta$.}.

Now, what we need to compute remains the $U(1)$-equivariant index of $D_{10}$, 
\beqa
&&\Tr_{\mathbb{X}}e^{tH}-\Tr_{\Xi}e^{tH}=
\Tr_{ker D_{10}}e^{tH}-\Tr_{Coker D_{10}}e^{tH}:={\rm ind} D_{10}.
\label{index}\eeqa
 To see this, note that the $D_{10}$ maps the eigenmode of $H$ on the bundle $\mathbb{X}$ to the eigen mode with the same eigenvalues on the $\Xi$, unless these modes are in kernel or cokernel of the operator $D_{10}$. 

To compute the index, (\ref{index}), we will first show that the operator $D_{10}$ is transversally elliptic with respect to the $U(1)$ action generated by $H$, i.e. elliptic in all directions transversal to the $H$-orbit.   If the operator is transversally elliptic on compact manifold, it is guaranteed that each subspace with the same eigenvalue of $H$ in kernal and cokernel is finite dimensional \cite{Atiyah1, Atiyah2}. We will assume that it still holds for the AdS$_2\times$S$^2$ and we will compute  the  index (\ref{index}) using Atiyah-Bott fixed point formula.  

To show the transversally ellipticity, we compute the symbol of the operator $D_{10}$. The $D_{10}$ appears as $\Xi D_{10}\mathbb{X}$ in the expression of $\cV$ in (\ref{formalQV}).  We take the relevant terms,
\be
 \frac{1}{(4\cosh\eta)^2}\left[ (\hat{\cQ}\Psi^{I\mu})^{\dagger}\Psi^I_{\mu}+\half (\hat{\cQ}\Xi^{Iij})^{\dagger}\Xi^{Iij}\right]+i\R^{-1}\bar{c}^I\nabla_{\mu}W^{I\mu}\,,
\ee
and use the explicit expression in (\ref{daggeronfields}) with neglecting $\hat{\cQ}c^I$ and $\hat{\cQ} \Xi^{Iij}$ as they are not relevant for the operator $D_{10}$.  To explicitly write the symbol of $D_{10}$ operator, denoted as $\sigma(D_{10})$, we consider only the highest derivative terms and replace $\partial_\mu$ by $ip_\mu$. It is convenient to introduce orthonormal four unit vector fields $u_{a}^{\mu}$ as,
\be\ba{rl}
i (\sigma^{a})_{i}{}^{j}\bar{\xi}_{j}\gamma^{\mu}\xi^{i}&=2\sqrt{\cosh^2(\eta)-\cos^2(\psi)}\,u_{a}^{\mu}\,,~~~\sigma^{a}: \mbox{ Pauli' sigma }, ~a=1,2,3\,,\\
\bar{\xi}_{i}\gamma^{\mu}\xi^{i}&=2\sqrt{\cosh^2(\eta)-\cos^2(\psi)}\,u_{4}^{\mu}\,.
\ea\ee
In particular,
\be
\cL_v=v^{\mu}\partial_{\mu}= 4\sqrt{\cosh^2(\eta)-\cos^2(\psi)}\,u^{\mu}_4 i p_\mu=4\sqrt{\cosh^2(\eta)-\cos^2(\psi)}\,i p_4\,.
\ee
We also define
\be
\Xi^{I}_a:= i\half \Xi^{I}_{i}{}^{j}(\sigma_a)_{j}{}^{i}=i\half\varepsilon_{ik}\Xi^{Ikj}(\sigma_a)_{j}{}^{i},
\ee
equivalently
\be
\Xi^{I}_a(\sigma_a)_{j}{}^{i}=i \varepsilon_{jk}\Xi^{Iki}\,.
\label{Xia}\ee
Then, the highest derivative terms of $\Xi D_{10}\mathbb{X}$ term are
\be
\frac{1}{2ch^2_{\eta}}\begin{pmatrix}\Xi^{I}_1\\ \Xi^{I}_{2}\\ \Xi^{I}_{3}\\ \bar{c}^{I}\\ c^{I}\end{pmatrix}^{T} \!\!\!\!\!\cdot\!\!
\begin{pmatrix}  c_{\psi}p_{4}&  ch_{\eta}p_{3}&-ch_{\eta}p_2&-c_{\psi}p_{1}&\alpha p_{1}
\\
-ch_{\eta}p_{3}&c_{\psi}p_{4}&ch_{\eta}p_{1}&-c_{\psi}p_2&\alpha p_{2}
\\
ch_{\eta}p_{2}&-ch_{\eta}p_{1}&c_{\psi}p_4&-c_{\psi}p_{3}&\alpha p_{3}
\\
-\frac{1}{\R}ch^2_{\eta}p_1&-\frac{1}{\R}ch^2_{\eta}p_2 &-\frac{1}{\R} ch^2_{\eta}p_3 &- \frac{1}{\R }ch^2_{\eta} p_4&0\\
\half  \alpha p_4p_1&\half \alpha p_4 p_2&\half \alpha p_4 p_3&\alpha (\half p_4p_4-p^a p_a) & -c_{\psi}p^ap_a\end{pmatrix}
\!\!\cdot\!\!\begin{pmatrix}W^{I}_{1}\\W^{I}_{2}\\W^{I}_{3}\\W^{I}_{4}\\X^{I}_2\end{pmatrix}\,,
\label{D10}\ee
\\
where we denoted
\be
c_{\psi}=\cos(\psi)\,,~~s_{\psi}=\sin(\psi)\,,~~ch_{\eta}=\cosh(\eta)\,,~~sh_{\eta}=\sinh(\eta)\,,~~
\alpha =\sqrt{ch^2_{\eta}-c^2_{\psi}}\,.
\ee
The matrix $\sigma(D_{10})$ can be block diagonalized by suitable change of variables within $\mathbb{X}$ and $\Xi$.  By changing
\be\ba{l}
W^{I}_4 \rightarrow \frac{c_\psi}{ch_{\eta}} W^{I}_4 - \frac{\alpha}{ch_{\eta}}  X^{I}_2\,,  ~~~~~X^{I}_2 \rightarrow \frac{\alpha}{ch_\eta}W^{I}_4 + \frac{c_{\psi}}{ch_\eta}X^{I}_2\,,\\
\bar{c}^{I}\rightarrow \frac{1}{\R}ch_{\eta}\,\bar{c}^{I}-\half \frac{\alpha}{ch_{\eta}}p_4 \,c^{I}\,,~~~~~
c^{I}\rightarrow \frac{1}{\R}\alpha\frac{ p_4}{ p^{a}p_{a}} \,\bar{c}^{I} +(1-\half \frac{\alpha^2}{ch_{\eta}^2}\frac{ p_4 p_4}{p^a p_a}) \,c^{I}\,,
\ea\ee
we get 
 \be
 \sigma(D_{10})=
\frac{1}{2ch^2_{\eta}}
\begin{pmatrix}  c_{\psi}p_{4}&  ch_{\eta}p_{3}&-ch_{\eta}p_2&-ch_{\eta}p_{1}&0
\\
-ch_{\eta}p_{3}&c_{\psi}p_{4}&ch_{\eta}p_{1}&-ch_{\eta}p_2&0
\\
ch_{\eta}p_{2}&-ch_{\eta}p_{1}&c_{\psi}p_4&-ch_{\eta}p_{3}&0
\\
-ch_{\eta}p_1&-ch_{\eta}p_2 & -ch_{\eta}p_3 &- c_{\psi}p_4&0\\
0&0&0&0 & {-ch_{\eta}p^ap_a}\end{pmatrix}\,.
\ee
Nontrivial contribution to the index arises from the upper-left $4\times 4$ block of the matrix in the middle,
\be
\sigma(D'_{10})= \begin{pmatrix}  c_{\psi}p_{4}&  ch_{\eta}p_{3}&-ch_{\eta}p_2&-ch_{\eta}p_{1}
\\
-ch_{\eta}p_{3}&c_{\psi}p_{4}&ch_{\eta}p_{1}&-ch_{\eta}p_2
\\
ch_{\eta}p_{2}&-ch_{\eta}p_{1}&c_{\psi}p_4&-ch_{\eta}p_{3}
\\
-ch_{\eta}p_1&-ch_{\eta}p_2 & -ch_{\eta}p_3 &- c_{\psi}p_4 \end{pmatrix}\,.
\ee
We note that the above matrix is not invertible at the equator $\cos\psi=0$ of the S$^2$. This is because $\sigma\sigma^{T}= (\cosh^2{\eta} (p_1^2 +p_2^2 +p_3^2)+ \cos^2{\psi}p_4^2)\cdot {\mathbb{I}}\,$ and is zero for $p_1=p_2=p_3=0$ and $p_4\neq 0$. 
However, if we restrict the momentum to be orthogonal to the Killing vector $v^{\mu}$, then $\sigma$ is invertible as long as $(p_1\,,p_2\,,p_3)$ are not all zero. Therefore the operator $D_{10}$ is transversally elliptic with respect to the symmetry $\cL_v$ .

Now, we use the Atiyah-Bott fixed point formula to compute the equivariant index (\ref{index}). The Atiyah-Bott formula is reviewed in the appendix \ref{ABformula} and the formula is give by
\be
{\rm{ind}}(D_{10})=\sum_{A}\frac{\Tr _{\mathbb{X}}(\gamma)-\Tr_{\Xi}(\gamma)}{\det (1-\partial f(x)/\partial x)}\,,~~~~~~A=\{x|f(x)=x\}\,,
\label{A-Bformula}\ee
where the $\gamma$ is the transformation of the section induced by the $f(x)$.
The formula reduces the trace of the operator $e^{tH}$ into the summation over the fixed point of the operator $H$. In our case, there are two fixed points. One is the north pole of the S$^2$ together with the origin of the AdS$_2$ and the other is the south pole of the S$^2$ together with the origin of the AdS$_2$. Near the fixed points the space is locally $\mathbb{R}^2 \times \mathbb{R}^2$ so it is parametrized by the orthonormal coordinate $(x^1\,,x^2\,,x^3\,,x^4)$, where the $(x^1\,,x^2)$ are the local coordinate on the AdS$_2$ and $(x^3\,,x^4)$ are the local coordinate on the S$^2$. Let us define the complexified coordinates 
\be\ba{l}
z^1:=x^1 +i x^2\,,~~~~~ z^2 := x^3+i x^4\,,~~~~~\mbox{at  north pole}\,,\\
w^1:=x^1 +i x^2\,,~~~~~ w^2 := x^3+i x^4\,,~~~~~\mbox{at south pole}\,.\\
\ea\ee
Under the operator $e^{tH}$ they transform as
\be\ba{ll}
z^1 \rightarrow e^{it/\R}z^1:= q z^1\,,~~~~& z^2 \rightarrow e^{-it/\R}z^2:= \bar{q} z^2\,,\\
w^1 \rightarrow e^{it/\R}w^1:=q w^1\,,~~~~& w^2 \rightarrow e^{it/\R}w^2:= q w^2\,.
\ea\label{coordtransf}\ee
Note here that the operator $H$ generates as the $L-J$ rotation, where  $L$ and $J$ are the rotation on the AdS$_2$ and S$^2$. So the the coordinate $z^2$ rotates in the opposite way to the $z^1$. Also, the coordinate $w^2$ rotates the opposite again to the $z^2$ coordinate as it is the coordinate at the south pole of the S$^2$.

Let us consider how the fields transform. The nontrivial part is for $\Xi^{ij}$. One can explicitly see from (\ref{D10}) that  the fermions $(\Xi_1\,,\Xi_2\,,\Xi_3)$ defined in (\ref{Xia}) are dual of self-dual field at the north pole and the dual of anti-self-dual field at south pole such that they are contracted with $(F^+_{14}\,,F^{+}_{13}\,,F^{+}_{12})$ and $(F^{-}_{14}\,,F^{-}_{13}\,,F^{-}_{12})$ respectively\footnote{In the case we choose another Killing spinor which squares to $L+J$, we get opposite transformation rule for the $z^2$ and $w^2$. But now $(\Xi_1\,,\Xi_2\,,\Xi_3)$ are dual of $(F^{-}_{14}\,,F^{-}_{13}\,,F^{-}_{12})$ at North pole and $(F^{+}_{14}\,,F^{+}_{13}\,,F^{+}_{12})$ at South pole respectively, so we get same transformation rule as in (\ref{Nfieldtransf}) and (\ref{Sfieldtransf}), giving the same result (\ref{indD10}). Notice that in the $S^4$ computation, for the similar killing spinor which squares to $L+J$, the fields $(\Xi_1\,,\Xi_2\,,\Xi_3)$ are dual of $(F^{+}_{14}\,,F^{+}_{13}\,,F^{+}_{12})$ at North pole and $(F^{-}_{14}\,,F^{-}_{13}\,,F^{-}_{12})$ at South pole respectively. This is the main difference of the index computation in $S^4$ and AdS$_2\times$ S$^2$.}. In terms of the complexified coordinates, the self-dual and anti-self-dual field strength have the following basis,
\be\ba{ll}
F^{+}_{12}\sim {\rm{d}}z^1 \wedge{\rm d}\bar{z}^1 + {\rm d}z^2\wedge {\rm d}\bar{z}^2\,,~~~~~~&F^{-}_{12}\sim {\rm d}w^1 \wedge{\rm d}  \bar{w}^1-{\rm d}w^2 \wedge {\rm d}\bar{w}^2\,,\\
F^{+}_{13}\sim  {\rm{d}}z^1 \wedge{\rm d}z^2 + {\rm d}\bar{z}^1\wedge {\rm d}\bar{z}^2\,,~~~~~~& F^{-}_{13}\sim  {\rm d}w^1 \wedge {\rm d}\bar{w}^2+{\rm d}\bar{w}^1 \wedge {\rm d}w^2\,,\\
F^{+}_{14}\sim  {\rm{d}}z^1 \wedge{\rm d}z^2 - {\rm d}\bar{z}^1\wedge {\rm d}\bar{z}^2\,,~~~~~~& F^{-}_{14}\sim  {\rm d}w^1 \wedge {\rm d}\bar{w}^2-{\rm d}\bar{w}^1 \wedge {\rm d}w^2\,.
\ea\ee
By the definition (\ref{Xia}), we see that the fermions $(\Xi^{11}\,,\Xi^{22}\,,\Xi^{12})$ have the following basis at north pole,
\be
\Xi^{11}\sim \frac{\partial}{\partial z^{1}}\wedge \frac{\partial}{\partial z^2}\,,~~~\Xi^{22}\sim \frac{\partial}{\partial \bar{z}^{1}}\wedge \frac{\partial}{\partial \bar{z}^2}\,,~~~\Xi^{12}\sim \frac{\partial}{\partial z^{1}}\wedge \frac{\partial}{\partial \bar{z}^1}+ \frac{\partial}{\partial z^{2}}\wedge \frac{\partial}{\partial \bar{z}^2}\,,
\ee
and at the south pole,
\be
\Xi^{11}\sim \frac{\partial}{\partial w^{1}}\wedge \frac{\partial}{\partial \bar{w}^2}\,,~~~\Xi^{22}\sim \frac{\partial}{\partial \bar{w}^{1}}\wedge \frac{\partial}{\partial w^2}\,,~~~\Xi^{12}\sim \frac{\partial}{\partial w^{1}}\wedge \frac{\partial}{\partial \bar{w}^1}- \frac{\partial}{\partial w^{2}}\wedge \frac{\partial}{\partial \bar{w}^2}\,.
\ee
We now spell the transformation of fields at the fixed points. At the north pole,
\beqa\label{Nfieldtransf}
&&\gamma[X_2]=1\,,~~\gamma[W_{z^1}]=q\,,~~\gamma[W_{\bar{z}^1}]=\bar{q}\,,~~\gamma[W_{{z}^2}]=\bar{q}\,,~~\gamma[W_{\bar{z}^2}]=q\,,\\
&&\gamma[\Xi^{11}]=q^{-1}\bar{q}^{-1}=1\,,~~\gamma[\Xi^{22}]=\bar{q}^{-1}q^{-1}=1\,,
~~\gamma[\Xi^{12}]=1\,,
~\gamma[c]=1\,,
~\gamma[\bar{c}]=1\,.\nn
\eeqa
and similarly we get at the south pole 
\beqa\label{Sfieldtransf}
&&\gamma[X_2]=1\,,~~\gamma[W_{w^1}]=q\,,~~\gamma[W_{\bar{w}^1}]=\bar{q}\,,~~\gamma[W_{{w}^2}]=q\,,~~\gamma[W_{\bar{w}^2}]=\bar{q}\,,\\
&&\gamma[\Xi^{11}]=q^{-1}\bar{q}^{-1}=1\,,~~\gamma[\Xi^{22}]{=\bar{q}^{-1}q^{-1}=1}\,,
~~\gamma[\Xi^{12}]=1\,,
~\gamma[c]=1\,,
~\gamma[\bar{c}]=1\,.\nn
\eeqa

Applying the fixed point formula (\ref{A-Bformula}) and using the (\ref{coordtransf}), (\ref{Nfieldtransf}) and (\ref{Sfieldtransf})\,, we obtain the  following result  for each vector multiplet,
\be
{\rm{ind}}D_{10}=\left[ \frac{2q}{(1-q)^2}\right]_{+}+\left[ \frac{2q}{(1-q)^2}\right]_{-}\,.
\label{indD10}\ee
The first term is from origin of AdS$_2$ and north pole of S$^2$ and the second term is from origin of AdS$_2$ and south pole of S$^2$.  One obtains the degeneracies of eigen values of $H$ by expanding this expression in power series of $q$. 
Here we follow the way of expansion as was done in \cite{Pestun:2007rz,Atiyah2}. The result for the determinant is in fact independent on the way of expansions for the index. By expanding 
\be
\left[\frac{1}{1-q}\right]_{+}= \sum_{n=0}^{\infty} q^{n}\,,~~~~~~~~~~~\left[\frac{1}{1-q}\right]_{-}= -q^{-1}\sum_{n=0}^{\infty} q^{-n}\,,
\ee
we finally arrive at 
\be
{\rm ind} D_{10}=\sum_{n=-\infty}^{\infty}|2n| q^n\,.
\label{indD10}\ee

From the result (\ref{indD10}), we read off the 1-loop partition function for $n_v +1$ vector multiplets, 
\be\label{oneloopZ}
Z_{1-loop}=\left(\frac{\det K_{f}}{\det K_{b}}\right)^{\frac{1}{2}(n_v+1)}= \prod_{n=1}^{\infty}\left(\frac{n}{\R}\right)^{-2n(n_v+1)}\,.
\ee
Using the $\zeta$-function regularization,
\be
\log Z_{1-loop}=-(n_v+1)\sum_{n=1}^{\infty} (2n) \log \R^{-1}= -\frac{n_v+1}{6}\log \R\,.
\ee
For given radius $\R$, we get the exact 1-loop partition function. However, we note that this radius $\R$ is not the physical radius because it can be chosen to be an arbitrary constant value as the choice of $D$-gauge fixing. In the next section, we will show that, by appropriate integration measure, the 1-loop partition function is independent of the gauge choice and depends on the radius of the physical AdS$_2\times$ S$^2$ metric.

\section{Integration measure}
In the previous section, we assumed that the integration measure is trivial. As a result, the one-loop partition function is not independent of the choice of $D$-gauge. This implies that the trivial path integration measure is not scale invariant.  In this section, we will properly define the path integration measure and show that the result of the $1$-loop partition function is indeed gauge invariant. Further, the result depends on the solutions of localization equations through the radius of the physical AdS$_2$$\times$S$^2$ metric. 
To define the measure we use the ultra locality arguments \cite{Polchinski:1985zf,Moore:1985ix}, as well as the condition that the result should be in terms of the physical quantities.

Let us consider the kinetic terms in the action,
\be
\int {\rm d}x^{4}\sqrt{g}\left[e^{-K}R_{g} +N_{IJ}\left( \partial_{\mu}X^I\partial_{\nu}\bar{X}^Jg^{\mu\nu}+F^I_{\mu\nu}F^J_{\lambda\rho}g^{\mu\lambda}g^{\nu\rho}+\bar{\lambda}^I\slashed{\partial}_{g}\lambda^J -Y^I_{ij}Y^{Jij}\right)\right]\,.
\ee 
The metric $g_{\mu\nu}$ is not a physical metric (it has dilatation weight -2) and is related to physical metric $G_{\mu\nu}$ which is the metric in Einstein frame by redefinition,
\be
G_{\mu\nu}=g_{\mu\nu}e^{-K}\,,~~~~~~~~~e^{-K}=\frac{\Rp^2}{\R^2} \,.
\ee
Note that the radius of the AdS$_2 \times$ S$^2$ metric $g_{\mu\nu}$ is fixed to the constant $\R$ and the physical radius $\Rp$ is not fixed but depends on the scalars, i.e. $\Rp= \Rp(X\,,\bar{X})$ as the K$\ddot{\text a}$hler potential $K$ is the function of the scalars. In terms of the physical metric, we get standard Einstein-Hilbert action, and the kinetic term of the vector multiplets fields are
\be
\int {\rm d}x^{4}\sqrt{G}N_{IJ}\left[e^{K}\partial_{\mu}X^I\partial_{\nu}\bar{X}^JG^{\mu\nu}+F^I_{\mu\nu}F^J_{\lambda\rho}G^{\mu\lambda}G^{\nu\rho}+e^{\frac{3}{2}K}\bar{\lambda}^I\slashed{\partial}_{G}\lambda^J-e^{2K}Y^I_{ij}Y^{Jij}\right]+\cdots\,.
\ee
Looking at the factors in front of the each kinetic term, the definition of  the norm for each field is defined as
\beqa
&&||\delta X||^2:=\int {\rm d}^4x\sqrt{G}e^{K}N_{IJ}  \delta X^I\delta\bar{X}^J=\int {\rm d}^4x\sqrt{g_0}\R^2 \Rp^2 N_{IJ}\delta X^I\delta\bar{X}^J\,, \nn\\
&&||\delta W||^2:=\int {\rm d}^4x\sqrt{G} \,N_{IJ} \delta W^I_{\mu}\delta W^J_{\nu}G^{\mu\nu}= \int {\rm d}^4x \sqrt{g_0} \Rp^2 N_{IJ}\delta W^I_{\mu}\delta W^J_{\nu}g_{0}^{\mu\nu}\,,\\
&&||\delta \lambda||^2 :=\int {\rm d}^4x\sqrt{G} e^{\frac{3}{2}K}N_{IJ}(\delta \lambda^{I}_{i}\delta \lambda^{Ji}\!+\!\delta \bar\lambda^{I}_{i}\delta \bar\lambda^{Ji})= \int {\rm d}^4x \sqrt{g_0}\R^{3} \Rp N_{IJ} (\delta \lambda^{I}_{i}\delta \lambda^{Ji}\!+\!\delta \bar\lambda^{I}_{i}\delta \bar\lambda^{Ji})\,,\nn\\
&&||\delta Y||^2 :=-\int {\rm d}^4x\sqrt{G} e^{{2}K}N_{IJ}\delta Y^I_{ij}\delta Y^{Jij}= -\int {\rm d}^4x \sqrt{g_0}\R^{4}  N_{IJ} \delta Y^I_{ij} \delta Y^{Jij}\,,\nn
\eeqa
where we denote $g_{0\mu\nu}$ as AdS$_2 \times$S$^2$ metric with unit radius.
By following the normalization conditions,
\be
1=\int \cD X\cD\bar{X}e^{-||\delta X||^2}=\int \cD W e^{-||\delta W||^2}=\int \cD\lambda \cD\bar{\lambda}e^{-||\delta \lambda ||^2}=\int \cD Y e^{-||\delta Y ||^2}\,,
\ee
the integration measure is determined as
\beqa
&&\cD X\cD\bar{X}= \prod_{x, I}{\rm d}X^{I}(x){\rm d}\bar{X}^{I}(x)\det( \Rp^2\R^2 N_{IJ})\,,\nn\\
&&\cD{W} =\prod_{x, I,\mu}{\rm d}W^{I}_{\mu}(x)\sqrt{\det \Rp^2 N_{IJ}}\,,\nn\\
&&\cD Y= \prod_{x,i,j,I}{\rm d}Y^{I}_{ij}(x) \sqrt{\det \R^{4} N_{IJ}}\,,\label{measure1}\\
&&\cD\lambda(x)\cD\bar{\lambda}(x)= \prod_{x, I, i}{\rm d}\lambda^{Ii}(x){\rm d}\bar{\lambda}^{Ii}(x)\det (\Rp \R^{3} N_{IJ})^{-1}\,.
\nn
\eeqa
Similarly, we determine the measure for the ghost multiplet by looking at the gauge fixing action in Einstein frame,
\beqa
&&\int {\rm d}x^{4}\sqrt{g}[\R^{-1}i\bar{c}^{I}\Box_{g}c^{I} +i \R^{-1}B^{I}\nabla_{\mu}W^{I\mu}+\R^{-2}\frac{\xi}{2}B^{I2}]\\
&&=\int {\rm d}x^{4}\sqrt{G}[e^{K}\R^{-1} i\bar{c}^I\Box_{G}c^I +i e^{2K}\R^{-1}B^I\nabla_{\mu}W^{I\mu}+e^{2K}\R^{-2}\frac{\xi}{2}B^{I2}+\cdots]\,.\nn
\eeqa
The definition of norm\footnote{We do not need $N_{IJ}$ because the gauge fixing action is chosen as (\ref{LGF}). }, 
\beqa
&&|| c||^2 :=\int {\rm d}^4x\sqrt{G} e^{K}\R^{-1} \bar{c}^I  c^{J}= \int {\rm d}^4x \sqrt{g_0}\Rp^{2}\R    \bar{c}^I c^{J}\\
&&||\delta B||^2 :=\int {\rm d}^4x\sqrt{G} e^{{2}K}\R^{-2}\delta B^I\delta B^{J}= \int {\rm d}^4x \sqrt{g_0}\R^{2}   \delta B^I \delta B^{J}\nn
\eeqa
and the normalization condition,
\be
1=\int \cD c \cD\bar{c}e^{-||c||^2}=\int \cD B e^{-||\delta B||^2}\,,
\ee
determine the integration measure for the ghost multiplets
\beqa
&&\cD c \cD \bar{c}=\prod_{x,I}{\rm d}c^I(x) {\rm d}\bar{c}^I(x)( \R \Rp^2 )^{-1} \,, \\
&&\cD B=\prod_{x,I}{\rm d}B^{I}(x)\R\,.\nn
\label{measure2}\eeqa
The measure (\ref{measure1}) and (\ref{measure2}) will give the result in terms of the physical quantities. One can consistently see that the 1-loop determinant for each kinetic operator for each field will be given in terms of
$
\det {\Box_{G}}$ and $\det {\slashed{\partial}}_{G}\,,
$
 not in terms of $
\det {\Box_{g}}$ and $\det {\slashed{\partial}}_{g}
$. 
It seems that a naive counting of the scale factor and radius factor, $\R$ and $\Rp$, bosonic measure and fermionic measure seems to be completely canceled. However, they are infinite product. The regularized number of those factors should not be canceled each other and should cancel the scale factor appears in the 1-loop partition function (\ref{oneloopZ}) such that the result should be only in terms of the physical radius $\Rp$.

Let us reconsider the computation of the 1-loop partition function. By the supersymmetric localization, the measure depends only on the saddle point value, i.e. $\Rp=\Rp(\vec{C})$, where $\vec{C}$ parametrizes all saddle point of the scalar in the vector multiplets.
We  now redefine the cohomological variables by following field redefinition,
\beqa\label{redefineX}
&&\tilde{\mathbb{X}}=(\tilde{X}_{2}\,,\tilde{W}_{\mu}):=(X_{2}\R\Rp(\vec{C})\,,W_{\mu}\Rp(\vec{C}))\,,\nn\\
&&\tilde{\Xi}=(\tilde{\Xi}^{ij}\,,\tilde{c}\,,\bar{\tilde{c}}):= (\Xi^{ij}\R^{3/2}\Rp(\vec{C})^{1/2}\,,c\,\R^{1/2}\Rp(\vec{C})^{3/2}\,,\bar{c}\,\R^{1/2}\Rp(\vec{C})^{1/2})\,.\eeqa
We also redefine the $\hat{Q}$ operator
\be
\tilde{\cQ}:=\R^{1/2}\Rp^{-1/2}(\vec{C})\hat{\cQ}\, .
\label{redefineQ}\ee
Since the $\Rp(\vec{C})$ is function of the saddle points,  $\tilde{\cQ}$ does not act on the $\Rp(\vec{C})$.
Using this operator we define the other primed cohomological variables,
\beqa\label{redefineQX}
&&\tilde{\cQ}\tilde{\mathbb{X}}:=(\tilde{\Psi}\,,\tilde{\Psi}_{\mu})=(\Psi \R^{3/2}\Rp(\vec{C})^{1/2}\,,  \Psi_{\mu}\R^{1/2}\Rp(\vec{C})^{1/2})\,,\nn\\
&&\tilde{ \cQ}\tilde{\Xi}:=(\tilde{B}^{ij}\,,-\tilde{\Phi}\,,\tilde{B})=(B^{ij}\R^2\,,-\hat{\Phi}\R\Rp(\vec{C})\,,B\R)\,.
\eeqa

In terms of these new variables and new supercharge, $\tilde{\mathbb{X}}\,, \tilde{\cQ}\tilde{\Xi}\,,\tilde{\cQ}\tilde{\mathbb{X}}\,, \tilde{\Xi}$,  we rewrite the localization lagrangian as in  (\ref{formalQV}),
\be
\sqrt{g}\hat{\cQ}\cV =\sqrt{g_0} \tilde{\cQ}\left[(\tilde{\cQ} \tilde{\mathbb{X}}'\,, \tilde{\Xi}')\begin{pmatrix}\tilde{D}_{00}&\tilde{D}_{01}\\ \tilde{D}_{10}&\tilde{D}_{11}\end{pmatrix}\begin{pmatrix}\tilde{\mathbb{X}}'\\\tilde{\cQ}\tilde{\Xi}' \end{pmatrix}\right]\,,
\label{formaltildeQV}\ee
where the $\tilde{D}_{ij}$ are properly defined by multiplying diagonal matrices whose elements are composed of $\R$ and $\Rp$.    We can follow the same analysis as below the (\ref{formalQV}) and arrive at computing the $U(1)$ equivariant index
\be
{\rm ind}\tilde{D}_{10}=\Tr_{\tilde{\mathbb{X}}} e^{t\tilde{H}}-\Tr_{\tilde{\Xi}} e^{t\tilde{H}}\,.
\ee
Here $\tilde{H}:=\tilde{\cQ}^2$ so the $\tilde{H}$ and $H$ are related by 
\be
\tilde{H}:=\frac{\R}{\Rp(\vec{C})}H\,. 
\ee
Therefore, the $q$ factor defined in (\ref{coordtransf}) is now replaced by
\be
q=e^{-it/\Rp}\,,
\ee
and we get the 1-loop partition function in terms of scale invariant length
\be
Z_{1-loop}=\prod_{n=1}^{\infty}\left(\frac{n}{\Rp(\vec{C})}\right)^{-2n(n_v +1)}= \exp\left[-{\frac{n_v +1}{6} \log(\Rp(\vec{C}))}\right]\,.
\label{Z1loop}\ee

\section{Conclusion and Discussion}
In this paper, we considered the $n_v+1$ $\cN=2$ vector multiplets on the AdS$_2\times$S$^2$ background and used the supersymmetric localization to compute their exact contribution to the quantum entropy function. We obtained the localization saddle point and computed the exact 1-loop partition function. In order to express the result in terms of physical radius of the AdS$_2\times$S$^{2}$, we proposed the scale invariant functional integration measure using the ultra locality argument. Collecting the result (\ref{Z1loop}) and the zero mode integral measure with the scale factors in (\ref{measure1}) or with the redefined fields as in (\ref{redefineX}), the functional integration of the quantum entropy function reduced to the following finite dimensional integration, 
\beqa
&W&= \int \prod_{I=1}^{n_v+1}\prod_{l\neq 0}{\rm d}\tilde{W}_{\mu}^{Il}{\rm d}\tilde{X}^{I0}_1{\rm d}\tilde{X}^{I0}_2{\rm d}\tilde{\lambda}^{Ii0}{\rm d}\bar{\tilde{\lambda}}^{Ii0}\, Z_{1-loop}\,e^{S_{ren}}\nn\\
&&=\int \prod_{I}^{n_v+1}\Rp^{-1}({{\rm d}C^I_1}\Rp)({\rm d}C^I_2\Rp)({\rm d}\vartheta^I{\rm d}\bar{\vartheta}^I\Rp^{-1})\, \Rp^{-\frac{1}{6}(n_v+1)}\,e^{S_{ren}}\nn\\
&&=\int {{\rm d}C^I_1}{\rm d}C^I_2{\rm d}\vartheta^I{\rm d}\bar{\vartheta}^I\, (\Rp(\vec{C}))^{-\frac{1}{6}(n_v+1)}\,e^{S_{ren}}\,.
\eeqa
Here, the $S_{ren}$ is the classical action on the localization manifold with the IR divergence removed. Since there is one zero mode for $X_2$, a pair of fermion zero mode appears as is argued in (\ref{Trace}). The $\vartheta\,,\bar{\vartheta}$ parametrize the fermion zero mode. The measure of each field has its own power of $\Rp$. Particularly, the infinite product of the boundary gauge modes integral gives us the regularized number of power, $\Rp^{-1}$ \cite{Banerjee:2010qc}.  Adding all the factors from measure and the 1-loop partition function, we result in the factor $\Rp^{-1/6}$ for each vector multiplet.

 The result explains the one of the key assumptions along the line to compute the exact quantum black hole entropy \cite{Dabholkar:2010uh,Dabholkar:2011ec,Gupta:2012cy}, where the classical measure was properly assumed in order to reproduce the result from the microstate counting.  We derived the contribution to the measure from the vector multiplets. Once computing all the 1-loop determinant for the hyper multiplets, gravitini multiplets and Weyl multiplet, one will be able to derive the measure for the $\cN=8$ supergravity (see the complementary work \cite{Murthy:2015yfa}). A slight difference from \cite{Dabholkar:2010uh,Dabholkar:2011ec,Gupta:2012cy} though is that  we get additional integration ${\rm d}C^I_2{\rm d}\vartheta^I{\rm d}\bar{\vartheta}^I$. If we chose the different choice of analytic continuation (\ref{Contour2}), the additional integration would not be appear. It is the artifact of the different choice of analytic continuation.  To be consistent, we expect that the fermion integration ${\rm d}\vartheta^I{\rm d}\bar{\vartheta}^I$ will cancel the contribution from the integration of ${\rm d}C^I_2$. While we left the explicit computation, one can see a consistency that the the $\Rp$ factor coming from the fermion zero mode cancels the power of $\Rp$ from the $X_2$ zero mode.  

The 1-loop determinant is scale invariant and depends on the physical metric.
It is remarkable that although we are considering abelian vector multiplets,  the 1-loop result depends on the continuous parameters $\vec{C}^I$ of localizing solutions through the physical metric. It is due to the proposed functional integral measure based on the fact that the measure should be scale invariant and the results in the one-loop determinant being dependent only on the physical quantities.  In fact, not only for the scale symmetry but also for supersymmetry, the measure should be invariant for the purpose of the localization. We were not able to show this and assumed that supersymmetric invariance is satisfied.

Our result for the one-loop partition function matches with the on-shell computation of the logarithmic correction in $\mathcal N=2$ black hole entropy \cite{Sen:2011ba}. In the on-shell computation the contribution of each vector multiplet to the logarithmic correction is $-\frac{1}{12}\ln A_H$ where $A_H=4\pi \Rp^2$ is the area of horizon. It is consistent that our measure factor $(\Rp(\vec{C}))^{-1/6}$ reproduces this logarithmic correction obtained from the on-shell computation.   The integration over $\vec{C}$ will not give further logarithmic correction as in \cite{Dabholkar:2011ec}. 

 For completing the story for the exact computation of the black hole entropy, we still have many open problems to solve. We have to incorporate the quantum fluctuation of all other multiplets, particularly Weyl multiplets. 
 In particular, including the hypermultiplets would be one of the tricky issues since there is no off-shell formulation of the hypermultiplets with finite number of auxiliary fields in supergravity, but it should be very important to treat the general $\cN=2$ supergravities because the hypermultiplets should be incorporated to complete the off-shell conformal supergravity as a compensating multiplets.  It will also help to have complete analysis of localization for the Weyl multiplets as mentioned in \cite{Gupta:2012cy}. Furthermore, one needs to also consider gravitini multiplets. It is particularly necessary for theories with higher supersymmetries like $\cN=4$ or $\cN=8$.   Here, we may have to understand similar issues that was appeared in our work on vector multiplets. 
For example, we may need to specify  the analytic continuations of all the fields, properly treat the gauge fixing of all the gauge symmetries in the conformal supergravity and should properly define the  gauge invariant functional measure for the path integral. We left these exercises for the future work.  




\section*{Acknowledgements} We wish to thank  Atish Dabholkar, Justin David, Naofumi Hama, Kazuo Hosomichi, Sameer Murthy, Tatsuma Nishioka, Tomoki Nosaka, Antoine Van Proeyen, Valentin Reys, Ashoke Sen, Seiji Terashima, Bernard de Wit for useful discussions. We would like thank Piljin Yi for suggesting this problem and helpful discussion.
\appendix
\section{Gamma matrices and spinors}\label{Gamma}
Our convention of gamma matrices and the reality properties follows the paper, \cite{VanProeyen:1999ni}.
\subsection{$(1,3)$ dimensions}\label{appendix1}
In Minkowskian four dimensions, there are two choices, $C_{\pm}$ and $B_{\pm}$, such that
\be\ba{llll}
\gamma_{a}^{\dagger}=-A\gamma_{a}A^{-1}\,,~&~A=\gamma_{0}\,,~&~A^{\dagger}=A^{-1}=-\gamma_{0}=-A\,,~&\\
\gamma_{a}^{T}=\mp C_{\pm}\gamma_{a}C_{\pm}^{-1}\,,~&~C_{\pm}^{T}=-C_{\pm}\,,~&~C_{\pm}^{\dagger}=C_{\pm}^{-1}\,,~&\\
\gamma_{a}^{*}=\pm B_{\pm}\gamma_{a}B_{\pm}^{-1}\,,~&~B_{\pm}^{T}=C_{\pm}A^{-1}\,,~&~B_{\pm}^{\dagger}=B_{\pm}^{-1}\,,~&~B_{\pm}^{*}B_{\pm}=\pm1\,.\\
\ea\label{Mgamma}\ee
Two representations are related by $C_{+}=C_{-}\gamma_{5}$.\\
Chirality operator
\be
\gamma_{5}=i\gamma_{0123}\,.
\ee
Useful relations
\be
(C_{\pm}\gamma_{1\cdots n})^{T}= -(-)^{n(n-1)/2}(\mp)^{n}C_{\pm}\gamma_{1\cdots n}\,.
\ee
\be
(C_{\pm}\gamma_{5})^{T}= -C_{\pm}\gamma_{5}\,.
\ee

The choice of $C_+$ and $B_+$ allows us to set Majorana spinor, defined as 
\be
\psi^{\dagger} A=\psi^{T}C_{+}\,,
\label{Majo}\ee
or equivalently,
\be
\psi^{*}= B_{+}\psi\,,
\ee
whereas,  for the choice of $C_-$ and $B_-$, the symplectic Majorana spinors can be defined\footnote{In general, $ -i\epsilon_{ij}$ may be replaced by arbitrary antisymmetric matrix satisfying $\Omega^{*}\Omega=-1$.},
\be
(\lambda^{i})^{*}=-i\epsilon_{ij}B_{-}\lambda^{j}\,,~~~~~~~\epsilon_{12}=\epsilon^{12}=1\,.
\ee
These spinors are not compatible with Weyl representation, such that under the chiral decompsition,
\be
(\psi_{\pm})^{*}= B_{+}\psi_{\mp}\,, 
\ee
\be
(\lambda_{\pm}^{i})^{*}=-i\epsilon_{ij}B_{-}\lambda_{\mp}^{j}\,.
\ee\label{WeylMajo}
Two chirally projected Majorana spinors $\psi_{\pm}^{i}$ symplectic Majorana spinors $\lambda_{\pm}^{i}$ can be related by
\be
\psi_{+}^{i}=\lambda_{+}^{i}\,,~~~~~~\psi_{-}^{i}=i \epsilon_{ij}\lambda_{-}^{j}\,.
\label{MajotoSMajo}\ee
\\
\subsection{$(0,4)$ dimensions}
In Euclidean four dimensions we also have two choices such that,
\be\ba{lll}
\gamma_{a}^{\dagger}=A\gamma_{a}A^{-1}\,,~&~A=1\,,~&~\\
\gamma_{a}^{*}=\gamma_{a}^{T}=\mp B_{\pm}\gamma_{a}B_{\pm}^{-1}\,,~&~B_{\pm}^{\dagger}=B_{\pm}^{-1}\,,~&~B^{T}_{\pm}=-B_{\pm}\,~\Leftrightarrow~B_{\pm}^{*}B_{\pm}=-1\,.\\
\ea\label{Egamma}\ee 
Since all gamma matrices are hermitian, the complex conjugation and the transpose are same.\\ 
Chirality operator
\be
\gamma_{5}^{E}=-\gamma_{1234}\,.
\ee
For two choices of $B_+$ and $B_-$, only symplectic Majorana-Weyl spinors can be defined, 
\be
(\rho^{i})^{*}=-i\epsilon_{ij}B_{\pm}\rho^{j}\,,
\ee
and compatible with Weyl condition,
\be
(\rho_{\pm}^{i})^{*}=-i\epsilon_{ij}B_{\pm}\rho_{\pm}^{j}\,.
\label{SMW}\ee

\subsection{Fierz identities}
It is useful to note the following gamma matrix algebra,
\be
\gamma_{a_{m}\cdots a_{1}}\gamma^{b_{1}\cdots b_{n}}=
\sum_{l=0}^{{\rm{min}}[m,n]}l!\binom{m}{l}\binom{n}{l}\gamma_{[a_{m}\cdots a_{l+1}}{}^{[b_{l+1}\cdots b_{n}}\delta_{a_{1}}^{~b_{1}}\cdots\delta_{a_{l}]}^{~b_{l}]}\,.
\ee
In particular for $\gamma_5=i\gamma_{0123}$,
\be\ba{l}
\gamma^{a_1 \cdots a_n}\gamma_{5}= i \frac{1}{(4-n)!}\varepsilon^{a_n \cdots a_1}{}_{b_{4-n}\cdots b_{4}}\gamma^{b_{4-n}\cdots b_{4}}\,,\\
\gamma_{5}\gamma^{a_1 \cdots a_n}= i \frac{1}{(4-n)!}\gamma^{b_{4-n}\cdots b_{4}}\varepsilon_{b_{4-n}\cdots b_{4}}{}^{a_n \cdots a_1},~~~~~~~\varepsilon^{0123}=1\,.\\
\ea\ee
For example
\be\ba{l}
\gamma^{a}\gamma_{5}=i\frac{1}{3}\varepsilon^{a b_{1}b_{2}b_{3}}\gamma_{b_{1}b_{2}b_{3}}\\
\gamma^{a_1 a_2}\gamma_{5}=i\frac{1}{2}\varepsilon^{a_2 a_1 b_{1}b_{2}}\gamma_{b_{1}b_{2}}\\
\gamma^{a_1 a_2 a_3}\gamma_{5}=i\varepsilon^{a_3 a_2 a_1 b}\gamma_{b}\\
\gamma_{abcd}\gamma_{5}=i \varepsilon_{abcd}\,.
\ea\ee
\be
\ee
Gamma matrices form a complete basis so it is followed by the Fierz arrangement,
\be\ba{l}
C_{\alpha\gamma}C_{\delta\beta}=\frac{1}{4}\sum_{n=0}^{4}\frac{1}{n!}(C\gamma_{a_{n}\cdots a_{a}})_{\delta\gamma}(C\gamma^{a_{1}\cdots a_{n}})_{\alpha\beta}\,,\\
(CP_{\pm})_{\alpha\gamma}(CP_{\pm})_{\delta\beta}=\half (CP_{\pm})_{\delta\gamma}(CP_{\pm})_{\alpha\beta}+\frac{1}{8}(C\gamma_{ba}P_{\pm})_{\delta\gamma}(C\gamma^{ab}P_{\pm})_{\alpha\beta}\,,\\
(CP_{\pm})_{\alpha\gamma}(CP_{\mp})_{\delta\beta}=\half (C\gamma_{a}P_{\pm})_{\delta\gamma}(C\gamma^{a}P_{\mp})_{\alpha\beta}\,,
\ea\ee
where $C$ can be either $C_+$ or $C_-$ in (\ref{Mgamma}).\\
For example, with bosonic fermions $\eta_{\pm}\,, \xi_{\pm}$ and $\lambda_{\pm}$ with positive or negative chiralities,
\be\ba{l}
\eta_{\pm}(\xi_{\pm}C\lambda_{\pm})=\half (\xi_{\pm}C\eta_{\pm})\lambda_{\pm}+\frac{1}{8}(\xi_{\pm}C\gamma_{ba}\eta_{\pm})\gamma^{ab}\lambda_{\pm}\,,\\
\eta_{\pm}(\xi_{\mp}C\lambda_{\mp})=\half (\xi_{\mp}C\gamma_{a}\eta_{\pm})\gamma^{a}\lambda_{\mp}\,.
\ea\ee

\subsection{Minkowskian theory to Euclidean theory for $\cN =2$}
While the Minkowskian space allows the Majorana representation, the Euclidean theory does not. They have different properties under the complex conjugation.
So in order to relate Minkowskian and Euclidean theory, we have to hide the complex conjugate operation. We change the Dirac conjugation of spinors, $\bar{\psi}:= \psi^{\dagger}A$, into the charge conjugation $\psi^{T}C_-\gamma_5$ using the Majorana relation given in (\ref{Majo}) and the relation between $C_+$ and $C_-$ in (\ref{Mgamma}).
Now, since the same $C_{-}$ can be used both Minkowskian and Euclidean spacetime, as in (\ref{Mgamma}) and (\ref{Egamma}),  Euclideanization is  straightforward.

In the case of $\cN=2$ theory, it is convenient to use the symplectic Majorana spinor representation because it is allowed both in Minkowskian and Euclidean theory. Using the relation (\ref{MajotoSMajo}), we can redefine spinor fields to satisfy the symplectic Majorana condition. After hidding the $\dagger$ operation in the theory using the symmplectic Majorana conjugate, one cannot distinguish whether it is Minkowskian or Euclidean theory and we are free to move by analytic continuation $t=-i\theta$.  

Tensor density that can be used in self or anti- selfdual equation should also be modified.
\be
e^{\mu\nu\lambda\rho}:=\frac{1}{\sqrt{-g}}\epsilon^{\mu\nu\lambda\rho}\,,~~~\epsilon^{t123}=1\,.
\label{densityM}
\ee 
In Euclidean space $t=-i\theta$
\be
e^{\mu\nu\lambda\rho}:=i\frac{1}{\sqrt{g}}\epsilon^{\mu\nu\lambda\rho}\,,~~~\epsilon^{\theta123}=1\,.
\label{densityE}\ee
In both of Minkowskian and Euclidean space,  the self or anti-selfdual condition is written in terms of the tensor density, $e^{\mu\nu\lambda\rho}$, in (\ref{densityM}) and (\ref{densityE}),
\be
T^{\mu\nu \pm}=\pm i\half e^{\mu\nu\lambda\rho} T_{\lambda \rho}^{~\pm}=\mp \textstyle{\frac{1}{\sqrt{g}}}\epsilon^{\mu\nu\lambda\rho} T_{\lambda \rho}^{~\pm}\,,
\label{selfdualE}\ee
where $T^{\mu\nu \pm}$ are self and anti-selfdual tensor respectively.

\section{Superconformal calculus for $d=4$ and $\cN =2$ SUGRA}\label{offsugra}
We review the superconformal calculus for $d=4$ and $\cN =2$ off-shell supergratives. We refer the reader to  \cite{Mohaupt:2000mj} for detailed review, and to  \cite{deRoo:1980mm, de Wit:1980tn, de Wit:1984px} for the original development. 
\subsection{Weyl multiplet}
The first step is to construct superconformal gauge theory by promoting all the $\cN=2$ superconformal generator as local symmetries.  By  all the local superconformal transformation,  the covariant derivative is defined as
\be
D_{\mu}:=\partial_{\mu}-\sum_{T}\delta(h_{\mu}(T))
\ee\label{covderiv}
where the sum is for all superconformal generators except the translation generator\cite{VanProeyen:1999ni}  and the $\delta$ is gauge transformation with the gauge field, $h_{\mu}(T)$, as parameter. Later, we will introduce $\cD_{\mu}$ as a covariant derivative with respect to $ M, D, A, V$. The gauge fields $h_{\mu}(T)$ and the symmetry parameters for each symmetry generators are contained in the table \ref{connection}, and the table \ref{Weyl} shows the charges of the gauge field and supersymmetry parameters. \\

{\scriptsize{\begin{table}[h]
\begin{center}
\begin{tabular}{c|cccccccc}
\hline
generator $T$~&~$P^{a}$&$M^{ab}$ &$D$&$K^{a}$&$Q^{i}$&$S^{i}$&$(V_{\Lambda})^{i}{}_{j}$&$A$ \\

Connection $h_{\mu}(T)$&~$e_{\mu}{}^{a}$&$\omega_{\mu}^{ab}$ &$b_{\mu}$&$f_{\mu}^{a}$&$\half\psi_{\mu}{}^{i}$&$\half \phi_{\mu}^{i}$&$-\half\cV_{\mu}{}^{i}{}_{j}$&$-iA_{\mu}$ \\
parameter &~$\xi^{a}$&$\varepsilon^{ab}$ &$\Lambda_{D}$&$\Lambda_{K}^{a}$&$\varepsilon^{i}$&$\eta^{i}$&$\Lambda_{V}{}^{i}{}_{j}$&$\Lambda_{A}$ \\
\hline

\end{tabular}
\caption{ Table of superconformal gauge fields and transformation parameters} 
\label{connection}
\end{center}
\end{table}}}

{\scriptsize{\begin{table}[h]
\begin{center}
\begin{tabular}{c|cccccccc|ccc|cc}
\hline
~&~$e_{\mu}{}^{a}$&$\psi_{\mu}^{i}$ &$b_{\mu}$&$A_{\mu}$&$\cV_{\mu}{}^{i}{}_{j}$&$T_{ab}^{ij}$&$\chi^{i}$&$D$ &$\omega_{\mu}^{ab}$&$f_{\mu}^{a}$&$\phi_{\mu}^{i}$&$\epsilon^{i}$&$\eta^{i}$\\
\hline
$\omega$&  $-1$&$-\half$&$0$&$0$&$0$&$1$&$\textstyle{\frac{3}{2}}$&0&0&1&$\half$&$-\half$&$\half$\\
$c$  &$0$&$-\half$&0&0&0&$-1$&$-\half$ &0&0&0&$-\half$&$-\half$&$-\half$ \\
$\gamma_{5}$&&$+$&&&&&$+$&&&&$-$&$+$&$-$\\
\hline
\end{tabular}
\caption{ Weyl weight $\omega$,  $U(1)_R$ weight $c$ and fermion chirality with respect to $\gamma_5$ for each the Weyl multiplet component field and supersymmetry parameters. } 
\label{Weyl}
\end{center}
\end{table}}}

The contents of the Weyl multiplet is given by the following $24+24$ off-shell degrees of freedom,
\be
(e_{\mu}{}^{a}\,, \psi_{\mu}^{i}\,, b_{\mu}\,, A_{\mu}\,, \cV_{\mu}{}^{i}{}_{j}\,,T_{ab}^{ij}\,,\chi^{i}\,,D)\,,
\ee
where the $\omega_{\mu}^{ab}\,,f_{\mu}^{a}\,, \phi_{\mu}^{i}$ are not included because they are not independent fields but composite fields in terms of others. The constrained relations are presented in (\ref{ConvConst}). The fields $T_{ab}^{ij}\,, \chi\,, D$ are the auxiliary tensor, spinor and scalar fields, and the auxiliary tenor satisfies
antiselfdual condition
\be
T_{ab}^{ij}=-\half i \epsilon_{abcd}T^{cdij}\,,~~~~~~~~\epsilon^{0123}=1\,,
\ee
whose complex conjugation gives selfdual tensor, 
\be
T_{abij}= (T_{ab}^{ij})^*\,.
\ee
Conventional notations are
\be\ba{l}
T_{ab}^{+}:= T_{abij}\varepsilon^{ij}\,, ~~~~~T_{ab}^{-}:=T_{ab}^{ij}\varepsilon_{ij}\,, ~~~~~\varepsilon_{ij}\varepsilon^{ij}=2
\\
T_{abij}=\half T_{ab}^{+}\varepsilon_{ij}\,,~~~~~T_{ab}^{ij}=\half T_{ab}^{-}\varepsilon^{ij}\,.
\ea\ee
The $SU(2)$ gauge fields $V_{\mu}{}^{i}{}_{j}$ is anti-hermitian and traceless
\be
\cV_{\mu}{}^{i}{}_{j}+ \cV_{\mu j }{}^{i}=0\,,  ~~~~~~~~\cV_{\mu}{}^{i}{}_{i}=0\,,~~~~~~~\mbox{where }~ \cV_{j}{}^{i}:=(\cV^{j}{}_{i})^{*}\,.
\ee
$\bullet${\bf Conventional constraints}:\\
In order to relate $\omega_{\mu}^{ab}\,,\phi_{\mu}^{i}\,,f_{\mu}^{i}$ with other fields, we impose the following constraints,
\be\ba{l}
R_{\mu\nu}(P)=0\,,\\
\gamma^{\mu}(\widehat{R}_{\mu\nu}(Q)^{i} +\sigma_{\mu\nu}\chi^{i})=0\,,~~~~~~~~~~~~~~~~~~~ \sigma_{\mu\nu}:=\half\gamma_{\mu\nu}\,,
\\
e_{b}{}^{\nu}\widehat{R}_{\mu\nu}(M)_{a}{}^{b}-i\widetilde{\widehat{R}}_{\mu a}(A)+\frac{1}{8}T_{abij}T_{\mu b}^{ij}-\frac{3}{2}D e_{\mu a}=0\,.
\label{ConvConst}\ea\ee
Here, the modified field strengths are
\be\ba{lll}
\widehat{R}_{\mu\nu}(Q)^{i}&=&2 \cD_{[\mu}\psi_{\nu]}^{i}-\gamma_{[\mu}\phi_{\nu ]}^{i}-\frac{1}{4}\sigma^{ab}T_{ab}^{ij}\gamma_{[\mu}\psi_{\nu ]j}
\\
\widehat{R}_{\mu\nu}(A)&=& 2\partial_{[\mu}A_{\nu]}-i \left( \half \bar{\psi}_{[\mu}^{i}\phi_{\nu]i} +\frac{3}{4}\bar{\psi}_{[\mu}^{i}\gamma_{\nu]}\chi_{i}-\mbox{h.c.}\right)
\\
\widehat{R}_{\mu\nu}(\cV)^{i}{}_{j}&=&2 \partial_{[\mu}\cV_{\nu]}{}^{i}{}_{j}+\cV_{[\mu}{}^{i}{}_{k}\cV_{\nu]}{}^{k}{}_{j}\\
&&+\left( 2 \bar{\psi}^{i}_{[\mu}\phi_{\nu]j}-3\bar{\psi}^{i}_{[\mu}\gamma_{\nu]}\chi_{j}-2 \bar{\psi}_{[\mu j}\phi_{\nu]}^{i}+3\bar{\psi}_{[\mu j}\gamma_{\nu]}\chi^{i}  \right)
\\
&&-\half \delta^{i}{}_{j}\left( 2 \bar{\psi}^{k}_{[\mu}\phi_{\nu]k}-3\bar{\psi}^{k}_{[\mu}\gamma_{\nu]}\chi_{k}-2 \bar{\psi}_{[\mu k}\phi_{\nu]}^{k}+3\bar{\psi}_{[\mu k}\gamma_{\nu]}\chi^{k}  \right)
\\
\widehat{R}_{\mu\nu}(M)^{ab}&=&2 \partial_{[\mu}\omega_{\nu]}^{ab}-2 \omega_{[\mu}^{ac}\omega_{\nu]}^{cb}-4 f_{[\mu}{}^{[a}e_{\nu]}{}^{b]}+(\bar{\psi}_{[\mu}^{i}\sigma^{ab}\phi_{\nu]i}+\mbox{h.c.})\\
&&\half \bar{\psi}_{[\mu}^{i}T^{ab}_{ij}\psi_{\nu]}^{j}-\frac{3}{2}\bar{\psi}_{[\mu}^{i}\gamma_{\nu]}\sigma^{ab}\chi^{i}-\bar{\psi}_{[\mu}\gamma_{\nu]}\widehat{R}^{ab}(Q)_{i} + \mbox{h.c.}
\ea\ee
and  the dual tensors are defined as,
\be
\widetilde{\widehat{R}}_{\mu\nu}(A)=\half i \epsilon_{\mu\nu\lambda\rho}\widehat{R}(A)^{\lambda\rho}\,.
\ee
Under the conventional constraints, (\ref{ConvConst}), the composite fields are expressed in terms of Weyl multiplet,
\be\ba{lll}
\omega_{\mu}^{ab}&=&-2 e^{\nu[a}\partial_{[\mu}e_{\nu]}{}^{b]}-e^{\nu[a}e^{b]\sigma}e_{\mu c}\partial_{\sigma} e_{\nu}{}^{c}-2 e_{\mu}{}^{[a} e^{b]\nu}b_{\nu}\\
&& -\frac{1}{4}(2 \bar{\psi}^{i}_{\mu}\gamma^{[a}\psi^{b]}_{i} + \bar{\psi}^{ai}\gamma_{\mu}\psi_{i}^{b}+ \mbox{h.c.})
\\
\phi^{i}_{\mu}&=& (\sigma^{\rho\sigma}\gamma_{\mu} -\frac{1}{3}\gamma_{\mu}\sigma^{\rho\sigma})(\cD_{\rho}\psi_{\sigma}^{i} - \frac{1}{8}\sigma^{ab}T_{ab}^{ij}\gamma_{\rho}\psi_{\sigma j} +\half \sigma_{\rho\sigma}\chi^{i})
\\
f_{\mu}{}^{i}&=& \half \widehat{R}_{\mu}{}^{a}-\frac{1}{4}(D+\frac{1}{3}\widehat{R})e_{\mu}{}^{a}-\half i \widetilde{R}_{\mu a}(A) +\frac{1}{16}T_{\mu b}^{ij}T_{ij}^{ab}\,,
\label{composite}\ea\ee
where
\be
\widehat{R}_{\mu}{}^{a}=\widehat{R}(M)_{\mu\nu}{}^{ab}e_{b}{}^{\nu}|_{f=0}\,,~~~~~~~~~\widehat{R}=\widehat{R}_{\mu}{}^{a}e_{a}{}^{\mu}\,.
\ee
\\~\\
$\bullet${\bf The transformation law and the superconformal algebra}\\
$Q-S-K-$ transformation rules for the Weyl multiplet fields are,
\be\ba{lll}
\delta e_{\mu}{}^{a}&=&\bar{\epsilon}^{i}\gamma^{a}\psi_{\mu i}+ \mbox{h.c.}\\
\delta \psi_{\mu}{}^{i}&=&2\cD_{\mu}\epsilon^{i}-\frac{1}{8}\gamma_{a}\gamma_{b}T^{abij}\gamma_{\mu}\epsilon_{j}-\gamma_{\mu}\eta^{i}\\
\delta b_{\mu}&=&\half \bar{\epsilon}^{i}\phi_{\mu i}-\frac{3}{4}\bar{\epsilon}^{i}\gamma_{\mu}\chi_{i}-\half \bar{\eta}^{i}\psi_{\mu i}+\mbox{h.c.} +\Lambda^{a}_{K}e_{\mu}{}^{a}\\
\delta A_{\mu}&=&\half i \bar{\epsilon}^{i}\phi_{\mu i} +\frac{3}{4}i\bar{\epsilon}^{i}\gamma_{\mu}\chi_{i}+\half i\bar{\eta}^{i}\psi_{\mu i}+\mbox{h.c.} \\
\delta \cV_{\mu}{}^{i}{}_{j}
&=&2 \bar{\epsilon}_{j}\phi_{\mu }{}^{i} -3 \bar{\epsilon}_{j}\gamma_{\mu}\chi^{i}+2 \bar{\eta}_{j}\psi_{\mu}{}^{i}
-2 \bar{\epsilon}^{i}\phi_{\mu j} +3 \bar{\epsilon}^{i}\gamma_{\mu}\chi_{j}-2 \bar{\eta}^{i}\psi_{\mu j}\\
&&-\half \delta^{i}{}_{j}(2 \bar{\epsilon}_{k}\phi_{\mu }{}^{k} -3 \bar{\epsilon}_{k}\gamma_{\mu}\chi^{k}+2 \bar{\eta}_{k}\psi_{\mu}{}^{k}
-2 \bar{\epsilon}^{k}\phi_{\mu k} +3 \bar{\epsilon}^{k}\gamma_{\mu}\chi_{k}-2 \bar{\eta}^{k}\psi_{\mu k})
\\
\delta T_{ab}^{ij}&=&8 \bar{\epsilon}^{[i}\widehat{R}_{ab}(Q)^{j]}\\
\delta\chi^{i}&=&-\frac{1}{12}\gamma_{a}\gamma_{b}\slashed{D}T^{abij}\epsilon_{j}+\frac{1}{6}\widehat{R}(\cV)^{i}{}_{j\mu\nu}\gamma^{\mu}\gamma^{\nu}\epsilon^{j}-\frac{1}{3}i \widehat{R}(A)_{\mu\nu}\gamma^{\mu}\gamma^{\nu}\epsilon^{i}\\
&&+D\epsilon^{i}+\frac{1}{12}T_{ab}^{ij}\gamma^{a}\gamma^{b}\eta_{j}
\\
\delta D&=& \bar{\epsilon}^{i}\slashed{D} \chi_{i} +\mbox{h.c.}\,,
\ea\ee
\be
\ba{lll}
\delta \omega_{\mu}{}^{ab}&=&-\bar{\epsilon}^{i}\sigma^{ab}\phi_{\mu i}-\half\bar{\epsilon}^{i}T^{ab}_{ij}\psi_{\mu}^{j}+\frac{3}{2}\bar{\epsilon}^{i}\gamma_{\mu}\sigma^{ab}\chi_{i}
\\&&+\bar{\epsilon}^{i}\gamma_{\mu}\widehat{R}^{ab}(Q)_{i}-\bar{\eta}^{i}\sigma^{ab}\psi_{\mu i}+ \mbox{h.c.}+ 2 \Lambda_{K}^{[a}e_{\mu}{}^{b]}
\\
\delta \phi_{\mu}{}^{i}&=&-2 f_{\mu}^{a}\gamma_{a}\epsilon^{i}-\frac{1}{4}\slashed{D}T_{cd}^{ij}\sigma^{cd}\gamma_{\mu}\epsilon_{j}+\frac{3}{2}\left[  (\bar{\chi}_{j}\gamma^{a}\epsilon^{j})\gamma_{a}\psi_{\mu}{}^{i}-(\bar{\chi}_{j}\gamma^{a}\psi_{\mu}{}^{j})\gamma_{a}\epsilon^{i}  \right]\\
&&+\half \widehat{R}(\cV)_{cd}{}^{i}{}_{j}\sigma^{cd}\gamma_{\mu}\epsilon^{j}+ i \widehat{R}(A)_{cd}\sigma^{cd}\gamma_{\mu}\epsilon^{i}+2 \cD _{\mu}\eta^{i}+ \Lambda_{K}^{a}\gamma_{a}\psi_{\mu}^{i}
\\
\delta f_{\mu}^{a}&=& -\half \bar{\epsilon}^{i}\psi_{\mu}^{j}D_{b}T^{ba}_{ij}-\frac{3}{4}e_{\mu}{}^{a}\bar{\epsilon}^{i}\slashed{D}\chi_{i}-\frac{3}{4}\bar{\epsilon}^{i}\gamma^{a}\psi_{\mu i} D\\
&& +\bar{\epsilon}^{i}\gamma_{\mu}D_{b}\widehat{R}^{ba}(Q)_{i}+\half \bar{\eta}^{i}\gamma^{a}\phi_{\mu i}+ \mbox{h.c.} +\cD_{\mu}\Lambda_{K}^{a}\,.
\ea\ee
\\
SUSY algebra
\be
[\delta_{Q}(\epsilon_{1}), \delta_{Q}(\epsilon_{2})]=\delta^{(cov)}(\xi)+\delta_{M}(\varepsilon)+\delta_{K}(\Lambda_{K})+\delta_{S}(\eta)+\delta_{gauge}\,,
\label{algebra}\ee
where
\be
\delta^{(cov)}(\xi):=\delta_{gct}(\xi)+\sum_{T}\delta_{T}(-\xi^{\mu}h_{\mu}(T))\,.
\ee
The sum over $T$ is for all superconformal transformation except the general coordinate transformation, and the parameters are\footnote{The sign convention for the Lorentz transformation is  $\delta_{M}(\varepsilon)e_{\mu}{}^{a}=-\varepsilon^{a}{}_{b}e_{\mu}{}^{b}$.}
\be\ba{lll}
\xi^{\mu}&=& 2\bar{\epsilon}_{2}^{i}\gamma^{\mu}\epsilon_{1i}+ \mbox{h.c.}
\\
\varepsilon^{ab}&=&\bar{\epsilon}_{2}^{i}\epsilon_{1}^{j}T_{ij}^{ab}+\mbox{h.c.}
\\
\Lambda_{K}^{a}&=&\bar{\epsilon}_{1}^{i}\epsilon_{2}^{j}D_{b}T_{ij}^{ab}-\frac{3}{2}\bar{\epsilon}_{2}^{i}\gamma^{a}\epsilon_{1i}D +\mbox{h.c.}
\\
\eta^i&=& 6\bar{\epsilon}^{i}_{[1}\epsilon^{j}_{2]}\chi_{j}\,,
\ea\ee
and the $\delta_{gauge}$  in general includes additional abelian, non-abelian or central charge gauge transformations.  

\be\ba{lll}
[\delta_{S}(\eta), \delta_{Q}(\epsilon)]&=&\delta_{M}\left(-2 \bar{\eta}^{i}\sigma^{ab}\epsilon_{i}+\mbox{h.c.}\right)+\delta_{D}\left(\bar{\eta}_{i}\epsilon^{i}+\mbox{h.c.}\right)+\delta_{A}\left(i\bar{\eta}_{i}\epsilon^{i}+\mbox{h.c.}\right)\\
&&+\delta_{V}\left(-2\bar{\eta}^{i}\epsilon_{j}+2\bar{\eta}_{j}\epsilon^{i}+\delta^{i}_{j}\bar{\eta}^{k}\epsilon_{k}-\delta^{i}_{j}\bar{\eta}_{k}\epsilon^{k} \right)\,,
\label{algebra2}\ea\ee

\be
[\delta_{S}(\eta_{1}), \delta_{S}(\eta_{2})]=\delta_{K}\left(\Lambda_{K}^{a}\right)\,,~~~\mbox{with }\Lambda_{K}^{a}=\bar{\eta}_{2i}\gamma^{a}\eta^{i}_{1}+\mbox{h.c.}\,.
\label{algebra3}\ee

\subsection{Vector multiplets}
Consider $N_{\upsilon}+1$ vector multiplet, restricting ourselves to the case of  abelian gauge symmetries,
\be
(X^{I}\,, \Omega_{i}^{I}\,,W_{\mu}^{I}\,, Y_{ij}^{I})\,, ~~~~~~~I=0,\cdots, N_{\upsilon}
\ee
Complex scalar $X^{I}$, a vector gauge field $W_{\mu}^{I}$\,, $SU(2)$ triplet auxiliary scalars $Y_{ij}^{I}$ 
\be
Y_{ij}^{I}=Y_{ji}^{I}\,, ~~ Y_{ij}=\varepsilon_{ik}\varepsilon_{jl}Y^{kl I}\,,
\ee
where $Y^{ij I}:= (Y_{ij}^{I})^*$.

{\scriptsize{\begin{table}[h]
\begin{center}
\begin{tabular}{c|cccc}
\hline
~&~$X^{I}$&$\Omega_{i}$ &$W_{\mu}^{I}$&$Y_{ij}^{I}$ \\
\hline
$\omega$&~$1$&$\textstyle{\frac{3}{2}}$ &$0$&$2$\\
$c$ &$-1$&$-\half$ &$0$&$0$\\
$\gamma_5$ &&$+$ &&\\

\hline

\end{tabular}
\caption{ Weyl weight $\omega$,  $U(1)_R$ weight $c$ and fermion chirality with respect to $\gamma_5$ for each vector multiplet component field } 
\label{vector}
\end{center}
\end{table}}}

One linear combination of the abelian gauge symmetries corresponds to the gauged central charge transformation, and the corresponding field strength belongs to the graviphoton. Note that we must have at least one vector multiplet in the theory in order to make contact with $\cN=2$ Poincar\'{e} supergravity, because the Weyl multiplet does not account for the graviphoton.\\
SUSY
\be\ba{lll}
\delta X^{I}&=& \bar{\epsilon}^{i}\Omega^{I}_{i}
\\
\delta \Omega_{i}^{I}&=&2 \slashed{D}X^{I}\epsilon_{i}+\half \varepsilon_{ij}\cF^{I \mu\nu -}\gamma_{\mu}\gamma_{\nu}\epsilon^{j}+ Y_{ij}^{I}\epsilon^{j}+ 2X^{I}\eta_{i}
\\
\delta W_{\mu}^{I}&=& \varepsilon^{ij}\bar{\epsilon}_{i}\gamma_{\mu}\Omega_{j}^{I}+ 2\varepsilon_{ij}\bar{\epsilon}^{i}\bar{X}^{I}\psi_{\mu}^{j}+\mbox{h.c.}
\\
\delta Y_{ij}^{I}&=& 2\bar{\epsilon}_{(i}\slashed{D}\Omega_{j)}^{I}+ 2 \varepsilon_{ik}\varepsilon_{jl}\bar{\epsilon}^{(k}\slashed{D}\Omega^{l)I}\,.
\label{vecsusy}\ea\ee
\be
\delta \cF^{I}_{ab}= -2 \varepsilon^{ij}\bar{\epsilon}_{i}\gamma_{[a}D_{b]}\Omega^{I}_{j}-2 \varepsilon^{ij}\bar{\eta}_{i}\sigma_{ab}\Omega_{j}^{I}+ \mbox{h.c.}
\ee
Here the covariant field strength is
\be
\cF^{I}_{\mu\nu}= F^{I}_{\mu\nu}-\left( \varepsilon_{ij} \bar{\psi}^{i}_{[\mu}\gamma_{\nu]}\Omega^{j I} + \varepsilon_{ij} \bar{X}^{I}\bar{\psi}^{i}_{\mu}\psi_{\nu}^{j}+\quarter \varepsilon_{ij}\bar{X}^{I}T^{ij}_{\mu\nu}+ \mbox{h.c.}\right)\,,
\label{CovariantF}\ee
which satisfies the Bianchi identity
\be
D^{b}\left(\cF^{+I}_{ab}  -\cF^{- I}_{ab}+\quarter X^{I}T_{abij}\varepsilon^{ij}-\quarter \bar{X}^{I}T^{ij}_{ab}\varepsilon_{ij}\right)={\textstyle\frac{3}{4}}\left( \bar{\chi}^{i}\gamma_{a}\Omega^{I j}\varepsilon_{ij}- \bar{\chi}_{i}\gamma_{a}\Omega^{I}_{j}\varepsilon^{ij} \right)\,,
\ee
where,
\be
\cF^{\pm I}_{ab}:=\half(\cF^{I}_{ab}\pm i\half \epsilon_{abcd}\cF^{Icd})\,.
\ee
The covariant derivatives are
\be\ba{l}
D_{\mu}X^{I}=\partial_{\mu}X^{I}-b_{\mu}X^{I}+iA_{\mu}X^{I}-\half \bar{\psi}^{i}_{\mu}\Omega_{i}^{I}\,,\\
D_{\mu}\Omega_{i}^{I}=(\partial_{\mu}+\quarter \omega_{\mu ab}\gamma^{ab}-\frac{3}{2} b_{\mu}+\half i A_{\mu})\Omega^{I}_{i}+\half \cV_{\mu i}{}^{j}\Omega_{j}^{I}\\
~~~~~~~~~~~~~~- \slashed{D}X^{I}\psi_{\mu i}-\quarter \varepsilon_{ij}\cF^{I ab -}\gamma_{a}\gamma_{b}\psi_{\mu}^{j}-\half Y_{ij}^{I}\psi_{\mu}^{j}-X^{I}\phi_{\mu i}\,.
\ea\ee
and the SUSY transformation of them are
\be\ba{l}
\delta (D_{a}X^{I})=\bar{\epsilon}^{i}D_{a}\Omega^{I}_{i}+\frac{3}{2}(\bar{\epsilon}_{i}\gamma_{a}\chi^{i})X^{I}-(\frac{1}{16}\bar{\epsilon}_{j}\gamma_{a}T_{bc}^{ji}\gamma^{bc}+\half \bar{\eta}^{i}\gamma_{a})\Omega_{i}^{I}\\
\delta (D_{a}\bar{X}^{I})=\bar{\epsilon}_{i}D_{a}\Omega^{iI}+\frac{3}{2}(\bar{\epsilon}^{i}\gamma_{a}\chi_{i})\bar{X}^{I}-(\frac{1}{16}\bar{\epsilon}^{j}\gamma_{a}T_{bcji}\gamma^{bc}+\half \bar{\eta}_{i}\gamma_{a})\Omega^{iI}\,.\\
\ea\ee
The algebra includes central charge gauge symmetry,
\be
\theta^{I}=4 \varepsilon^{ij}\bar{\epsilon}_{2i}\epsilon_{1j}X^{I}+ \mbox{h.c.}
\ee

Prepotential, $F(X)$ is a holomorphic function, which is homogeneous of secondegree, i.e.,  
\be
F(\lambda X)=\lambda^2 F(X)\,,
\ee
for any complex parameter $\lambda$. Some identities are
\be
F(X)=\half F_{I}X^{I}\,,~~~~F_{I}=F_{IJ}X^{J}\,,~~~~F_{IJK}X^{K}=0\,.
\ee
K$\ddot{\text a}$hler potential
\be
K=i(\bar{F}_{I}X^{I}-F_{I}\bar{X}^{I})=N_{IJ}X^{I}\bar{X}^{J}
\ee
Metric
\be
N_{IJ}=\partial_{I}\bar{\partial}_{J}K=-i(F_{IJ}-\bar{F}_{IJ})=2 {\rm{Im}} F_{IJ}\,.
\ee
Lagrangian(Bosonic):
\be\ba{ll}
e^{-1}\cL\sim&[i\bar{F}_{I}X^{I}(\frac{1}{6}R-D)+i\cD_{\mu}F_{I}\cD^{\mu}\bar{X}^{I}\\
&\,+\frac{1}{4}iF_{IJ}(F^{-I}_{ab}-\frac{1}{4}\bar{X}^{I}T_{ab}^{ij}\varepsilon_{ij})(F^{-J}_{ab}-\frac{1}{4}\bar{X}^{J}T^{ij}_{ab}\varepsilon_{ij})-\frac{1}{8}iF_{I}(F^{+I}_{ab}-\frac{1}{4}X^{I}T_{abij}\varepsilon^{ij})T^{ab}_{ij}\varepsilon^{ij}\\
&\,-\frac{1}{8}i F_{IJ}Y^{I}_{ij}Y^{Jij}-\frac{1}{32}i F(T_{abij}\varepsilon^{ij})^{2}] + \mbox{h.c.}
\label{veclagrangian}\ea\ee

Conventional gauge fixing conditions:
\be\ba{lc}
\mbox{$K$-gauge:}&~~b_{\mu}=0\,,\\
\mbox{$D$-gauge:}&~~-i(X^{I}\bar{F}_{I}-F_{I}\bar{X}^{I})=1\,,\\
\mbox{$U(1)$-gauge:}&~~X^{0}=\bar{X}^{0}\,.\\
\label{conventionalgauge}\ea\ee

\subsection{Chiral notation}
In the Minkowskian $\cN=2$ supereravities, we adopt the so-called chiral notation, which is to keep track of spinor chiralities through writing the $SU(2)_R$ index as an upper or lower index.
For instance, consider two Majorana spinors, $\psi^{i}$. The chiral projection of them are
\be
\Psi^{i}:=\half(1+\gamma_{5})\psi^i\,,~~~~~~~~\Psi_{i}:=\half(1-\gamma_{5})\psi^i\,.
\ee
Depending on the spinor an upper index might be associated with left or with right chirality. The assignments are listed in various tables  \ref{connection}, \ref{Weyl}, \ref{vector}. This chiral decomposition is not compatible with Majorana condition as we also see in (\ref{WeylMajo}), 
\be
(\Psi_{i})^{*}=B_{+}\Psi^{i}\,, ~~~~~~~\mbox{or equivalently }~(\Psi_{i})^{\dagger}A = \Psi^{i}C_{+}\,.
\ee
Since we can take $B_+=1$, the complex conjugation can be thought as  raising and lowering the $SU(2)_R$ indices. 

Dirac conjugation is defined as
\be
\bar{\Psi}^{i}:=(\Psi_{i})^{\dagger}A = \Psi^{i}C_{+}\,.
\ee
Note that
\be
\bar{\Psi}^{i}\gamma_{5}=\bar{\Psi}^{i}\,,~~~~\bar{\Psi}_{i}\gamma_{5}=-\bar{\Psi}_{i}\,,
\ee
which is followed by
\be\ba{l}
\bar{\Psi}_1^{i}\gamma_{a_{1}\cdots a_{n}}\Psi_2^{j}=0\,,~~~\mbox{for odd $n$}\\
\bar{\Psi}_1^{i}\gamma_{a_{1}\cdots a_{n}}\Psi_{2j}=0\,,~~~\mbox{for even $n$}\,.
\ea
\ee
We have further useful relations,
\be
\bar{\Psi}_1^{i}\gamma_{a_{1}\cdots a_{n}}\Psi_{2j}=(-1)^{\frac{1}{2} n(n+1)}\bar{\Psi}_{2j}\gamma_{a_{1}\cdots a_{n}}\Psi_{1}^{i}
\ee
\be
(\bar{\Psi}_1^{i}\gamma_{a_{1}\cdots a_{n}}\Psi_{2j})^{\dagger}=(-1)^{\frac{1}{2} n(n+1)}\bar{\Psi}_{2}^{j}\gamma_{a_{1}\cdots a_{n}}\Psi_{1i}=\bar{\Psi}_{1i}\gamma_{a_{1}\cdots a_{n}}\Psi_{2}^{j}\,.
\ee
\section{Bispinors}
We presents explicit values of some bispinors, which are useful for detailed calculation. 
\beqa
&&T_{ab}\xi_{1}\gamma^{ab}\xi^{1}=-T_{ab}\xi_{2}\gamma^{ab}\xi^{2}=-16 i (1+ \cos\psi \cosh\eta)\,,\nn\\&&T_{ab}\xi_{1}\gamma^{ab}\xi^{2}=T_{ab}\xi_{2}\gamma^{ab}\xi^{1}=16 i \sin\psi \sinh\eta\,,\nn\\
&&\bar{T}_{ab}\bar{\xi}_{1}\gamma^{ab}\bar{\xi}^{1}=-\bar{T}_{ab}\bar{\xi}_{2}\gamma^{ab}\bar{\xi}^{2}=16i (-1+ \cos\psi \cosh\eta)\,,\\
&&\bar{T}_{ab}\bar{\xi}_{1}\gamma^{ab}\bar{\xi}^{2}=\bar{T}_{ab}\bar{\xi}_{2}\gamma^{ab}\bar{\xi}^{1}=-16i\sin\psi\sinh\eta\,.\nn
\eeqa
To obtain (\ref{D10}), we note that
\beqa
&&i(\sigma^{1})_{i}{}^{j} (\xi_{j}\gamma^{mn}\xi^{i}+\bar{\xi}_{j}\gamma^{mn}\bar{\xi}^{i})F_{mn}=-8\cos\psi F_{14} -8\cosh\eta F_{23}\,,\nn\\
&&i(\sigma^{2})_{i}{}^{j} (\xi_{j}\gamma^{mn}\xi^{i}+\bar{\xi}_{j}\gamma^{mn}\bar{\xi}^{i})F_{mn}=8\cosh\eta F_{13} -8\cos\psi F_{24}\,,\\
&&i(\sigma^{3})_{i}{}^{j} (\xi_{j}\gamma^{mn}\xi^{i}+\bar{\xi}_{j}\gamma^{mn}\bar{\xi}^{i})F_{mn}=-8\cosh\eta F_{12} -8\cos\psi F_{34}\,.\nn
\eeqa

\section{Solution of localization equations for $X_2$}\label{X2sol}
In this section, we discuss solutions for $X_2$ to the localization equations. Our conjecture is that the normalizable regular solution is uniquely given as 
 (\ref{solutionX2}).  As evidence, we find asymptotic solutions and show that all possible solutions except (\ref{solutionX2}) that correspond to the asymptotic solutions may diverge at $r={\rm cosh}\, \eta=1$.
 
Two of localization equations from (\ref{bosonicL}) are $F^{I+}_{ab}=\frac{1}{8}X_2^{I} T_{ab}-\frac{1}{\xi_{j}\xi^{j}}v_{[a}D_{b]+}X_2^{I}$ and
$F_{ab}^{I-}=-\frac{1}{8}X_2^{I} \bar{T}_{ab}+\frac{1}{\bar{\xi}_{j}\bar{\xi}^{j}}v_{[a}D_{b]-}X_2^{I}$.
Substituting those equations into the Bianchi identity
$0=\partial_\theta F_{\eta\psi}+\partial_\psi F_{\theta\eta}+\partial_\eta F_{\psi\theta}$, we obtain the following differential equation for $X_2$,
\begin{align}
&-(\partial_\eta^2+\partial_\psi^2) X_2\ \frac{{\rm sinh}\,\eta \, {\rm cosh}\,\eta \, \sin\psi }{{\rm sinh}^2\eta\, + \sin^2\psi} \nonumber\\
&+\partial_\eta X_2\ \sin\psi \ \Big[ 
\frac{
 2 \cos^2\psi - {\rm cosh}^2 \eta
 }{
{\rm sinh}^2\eta\, + \sin^2\psi
 }  
 +
 2
 \frac{
 {\rm sinh}^2\eta \, {\rm cosh}^2\eta
 -
 \sin^2\psi\, \cos^2\psi
 }{
({\rm sinh}^2\eta\, + \sin^2\psi)^2
 }  
    \Big] \nonumber\\
    &+\partial_\psi X_2\ \cos\psi \ \Big[ 
 4
 \frac{
 \sin^2\psi \, 
 {\rm sinh}\, \eta \, {\rm cosh}\, \eta
 }{
({\rm sinh}^2\eta\, + \sin^2\psi)^2
 }  
-\frac{
 {\rm sinh}\, \eta \,  {\rm cosh}\, \eta
 }{
{\rm sinh}^2\eta\, + \sin^2\psi
 }  
 \Big] \nonumber\\
 &-\partial_{\theta}^2 X_2 \ \frac{{\rm cosh}\, \eta}{ {\rm sinh}\, \eta \, \sin \psi } \ =0 \, .
\end{align}
The asymptotic solutions to the above equation at $\eta\rightarrow\infty$ are
$e^{im\theta}\, Y_{\ell,m}(\psi,\phi)/{\rm cosh}^{\ell} \eta$ up to a multiplicative constant, where
$Y_{\ell,m}$ are spherical harmonics on $S^2$ and $\ell, m$ are non negative integer and $r={\rm cosh}\, \eta$.

If we multiply the above differential equation with ${\rm sinh}\,\eta \, \sin \psi \, ({\rm sinh}^2 \eta + \sin^2 \psi)^2$ and use the variable $r={\rm cosh}\, \eta$ and $x=\cos\psi$, we obtain the following equation
\begin{align}
&- (r^2-1) r (1-x^2) (r^2-x^2) \, \Big[ (r^2-1) \partial_r^2  + r \partial_r +(1-x^2) \partial_x^2-x \partial_x \Big] X_2 \nonumber\\
&+(r^2-1)(1-x^2)\, \Big[ (2x^2-r^2)(r^2-x^2) + 2((r^2-1)r^2-(1-x^2)x^2)  \Big] \partial_r X_2   \nonumber\\
&+(x^2-1) x \, \Big[4(1-x^2)(r^2-1)r-(r^2-1)r(r^2-x^2) \Big]\, \partial_x X_2 \nonumber\\
& - r (r^2-x^2)^2 \, \partial_{\theta}^2 X_2\ =0\, .
\end{align}

First let us restrict to solutions that are independent of $\theta$ and $\phi$. In this case, the asymptotic behavior is $Y_{\ell,0}(\psi,\phi)/r^{\ell}=P_{\ell}(x)/r^{\ell}$, where $P_{\ell}$ is Legendre polynomial. Since all the coefficient of the differential operators are polynomial of $r$ and $x$, we assume that the solutions can be written as
\begin{align}
X_2=\frac{P_{\ell}(x)}{r^{\ell}}+\sum_{n=\ell+1}^{\infty} \, \sum_{p=0}^{\infty} \frac{c_{n,p}\, x^p}{r^{n}} \, .
\end{align}

In case where $\ell=1$, the solution $X_2=x/r$ is a solution that is presented in (\ref{solutionX2}). In case where $\ell=2$, the simplest solution is
\begin{align}
X_2=\sum_{n=0}^{\infty} \big( \frac{3}{4n+2}x^2 -   \frac{3}{4n+6} \big)\frac{1}{r^{2n+2}} \, ,
\end{align}
and it diverges at $r=1$. We can rewrite it as follows
\begin{align}
X_2=\frac{1}{4r}\Big[ 6r+(r^2-x^2) {\rm log} \frac{r-1}{r+1}\Big] \,.
\end{align}
Although there are other solutions with $\ell=2$, if we fix the coefficient of the highest power and subtract the above solution from other solutions, the highest power of $r$ becomes less than $-2$. So we can consider them as solutions with $\ell>2$.

In case where $\ell=3$, the simplest solution is
\begin{align}
X_2=\sum_{n=0}^{\infty} \big( \frac{15}{4n+6}x^3 -   \frac{15}{4n+10}x \big)\frac{1}{r^{2n+3}}\, ,
\end{align}
which diverges at $r=1$ again. It can be rewritten as 
\begin{align}
X_2=\frac{5x}{2r} \, \Big[ 1+3(r^2-x^2)(1+\frac{r}{2}\, {\rm log} \frac{r-1}{r+1}) \Big] \, .
\end{align}
We conjecture that for any positive integer $\ell$, there is a unique solution of the following form
\begin{align}
X_2=\frac{P_{\ell}(x)}{r^{\ell}}+\sum_{k=1}^{\infty} \, \sum_{p=0}^{\lfloor \frac{\ell}{2} \rfloor} \frac{c_{\ell+2k,2p+e}\, x^{2p+e}}{r^{\ell+2k}} \, ,
\end{align}
where $e$ is $0$ or $1$ when $\ell$ is even or odd respectively and the floor function $\lfloor \frac{\ell}{2} \rfloor$ denotes the largest integer not greater than $\frac{\ell}{2}$. The series expansion of $r$ continues infinitely for $\ell \geq 2$.

Next let us consider solutions whose asymptotic behaviors are $e^{im\theta}Y_{\ell,\ell}(\psi,\phi)/r^{\ell}$. We propose that
\begin{align}
X_2=e^{i\ell \theta}\, \frac{Y_{\ell,\ell}(\psi,\phi)}{{\rm sinh}^{\ell}\, \eta}
\end{align}
are the solutions. They also diverge at $\eta=0$.
\section{Fixed point formula}\label{ABformula}
In this appendix we will show a proof of Atiyah-Bott fixed point formula \cite{Atiyah1, Atiyah2}, which we used in the section \ref{1-loop}.
Let  $E^0 \rightarrow E^1$ be an complex of vector bundles over a manifold $X$ and $D_{10}:\Gamma(E^0)\rightarrow \Gamma(E^1)$ is a differential operator. For a given map $f:X\rightarrow X$, we can define $f^{\ast}E^i$, which is a pullback of $E^i$ by $f$. Moreover the map $f$ induces a map $\gamma:E^i_{f(p)}\rightarrow E^i_{p}$\, , where $p\in X$. 

Let us define a map $T:=\gamma
 \circ
  f^{\ast}$. If the fixed points on $X$ under $f$ are isolated, we have the following formula
\be
\label{appD:AB}
{\rm Tr}_{{\rm Ker}\, D_{10}} T - {\rm Tr}_{{\rm Coker}\, D_{10}} T = \sum_{x\in {\rm fixed\, point\, set}} \frac{
{\rm Tr}_{E^0_x} \gamma-{\rm Tr}_{E^1_x} \gamma
}{\ |{\rm det}_{T_x X}(1-df(x))|\ }\, ,
\ee
if the left-hand side is well-defined. This is called Atiyah-Bott formula.

Since we focus on cases where $D_{10}$ commutes with $T$ here, ${\rm Tr}_{{\rm Ker}\, D_{10}}\, T - {\rm Tr}_{{\rm Coker}\, D_{10}}\, T={\rm Tr}_{\Gamma(E^0)} \,T - {\rm Tr}_{\Gamma(E^1)}\, T\, $.  Let us take an example where $E^0$ is an cotangent bundle and prove the following part in the formula\footnote{The quotation mark $``="$ is to emphasize that this equality holds only when the left hand side is well-defined. 
The left hand side is well-defined only when we take the difference of traces $\Tr_{\Gamma(E^0)} - \Tr_{\Gamma(E^1)}$. If the equality were strictly true, there would be no need to check transversally ellipticity of the operator, $D_{10}$.}
\be
\label{appD:AB2}
{\rm Tr}_{\Gamma(E^0)}\, T \ ``=" \sum_{x\in {\rm fixed\, point\, set}} \frac{
{\rm Tr}_{E^0_x} \gamma
}{\ |{\rm det}_{T_x X}(1-df(x))|\ } \, ,
\ee
in that case. The above equality holds if the left-hand side is well-defined. Since $\Gamma(E^0)$ is infinite dimensional, one should be careful about it.

 An element of $\Gamma(E^0)$ can be written as $\sum_{\mu}A_{\mu}(x)dx^{\mu}$. If $x$ goes to $y=f(x)$ by $f$, $A_{\mu}(x)dx^{\mu}$ is mapped to $A_{\mu}(y)dy^{\mu}$ by $f^{\ast}$ and it is further mapped to $A_{\nu}(y)(dy^{\nu}/dx^{\mu}) dx^{\mu}$ by $\gamma$. Generally if a given operator $Q$ maps $u_{\mu}(x) dx^{\mu}$ to $v_{\mu}(x)dx^{\mu} =: (Q^{\nu}_{\ \mu} u_{\nu}) dx^{\mu}$, 
we can define the kernel $K_{Q^{\nu}_{\ \mu}}$ for the operator 
$Q$ such that $\int dy \, K_{Q^{\nu}_{\ \mu}}(x,y) u_{\nu}(y)=v_{\mu}(x)$. In our case, the kernel for $T$ is
 \be
 \label{appD:KP}
 K_{T^{\nu}_{\ \mu}}(x,y)= \delta (f(x)-y) \, \frac{ dy^{\nu}}{d(f^{-1}(y))^{\mu}}\, .
\ee
For a general operator $Q$, the trace of $Q$ over $\Gamma(E^0)$ can be rewritten as follows
\be
\label{appD:kernel}
{\rm Tr}_{\Gamma(E^0)}\, Q \ ``=" \int dx\, \sum_{\mu}K_{Q^{\mu}_{\ \mu}}(x,x)\, ,
\ee
if the left-hand side is well-defined.
Let us derive the above relation. Fist we choose a complete set of orthonormal basis $\{ |A^{\{p\}}\rangle\}_{\{p\}}$ in $\Gamma(E^0)$\,.
Each component $A_{\mu}$ of one-form field $\sum_{\mu} A_{\mu}dx^{\mu}$ can take different field configuration labeled by $p_\mu$ and ${\{p\}}$ is the set $\{p_{\mu}\}_{\mu}$ of labels for all components. We can rewrite the trace in the left-hand side of the above relation as 
\be
\label{appD:5}
\sum_{\{p\}}\langle A^{\{p\}}| Q\,  A^{\{p\}}\rangle = \sum_{\{p\}} \int dx\, \sum_{\mu} (A^{\{p\}\, \mu}(x))^{\ast} \int dy \sum_{\nu} \, K_{Q^{\nu}_{\ \mu}}{(x,y)} A^{\{p\}}_{\nu}(y) \, ,
\ee
where we used the definition of the kernel $K_{Q^{\nu}_{\ \mu}}$.
By applying the following completeness condition
\begin{eqnarray}
\label{appD:pxy}
\sum_{\{p\}}\,  
(A^{\{p\}\, \mu}(x))^{\ast} A^{\{p\}}_{\nu}(y)=\delta (x-y) \delta^{\mu}_{\ \nu} \, 
\end{eqnarray}
to the right-hand side of (\ref{appD:5}), we can derive (\ref{appD:kernel}).

Let us apply the relation (\ref{appD:kernel}) to the operator $T$. By using (\ref{appD:KP}) and replacing the integration variable $x$ with $z:=x-f(x)$, we can derive the formula (\ref{appD:AB2}) as follows
 \begin{eqnarray}
 \displaystyle
{\rm Tr}_{\Gamma(E^0)}\, T&``="&\int dx\, \sum_{\mu}K_{T^{\mu}_{\ \mu}}(x,x)
=\int dx\, \sum_{\mu}
\delta(f(x)-x)  \frac{dx^{\mu}}{d(f^{-1}(x))^{\mu}} \nonumber\\
&=&\int dz \, |\frac{dx}{dz}|\, \delta(z) \sum_{\mu} \frac{dx^{\mu}(z)}{d(f^{-1}(x(z)))^{\mu}} \nonumber\\
&=&\int dz \, \frac{\delta(z) \sum_{\mu} \frac{dx^{\mu}}{d(f^{-1}(x))^{\mu})}}{|{\rm det}(1-\frac{df}{dx})|}   = \sum_{x \ \  {\rm s.t. } \, x=f(x)} \frac{ {\rm Tr}_{E^0_x} \gamma}{|{\rm det}(1-\frac{df}{dx})|}\, .
\end{eqnarray}
For general $E^i$, we can also derive the formula (\ref{appD:AB2}) in the same way. Although ${\rm Tr}_{\Gamma(E^i)}\, T$ itself may not be well-defined, ${\rm Tr}_{\Gamma(E^0)}\, T-{\rm Tr}_{\Gamma(E^1)}\, T$ is well-defined in the following cases.
If $T$ commutes with $D_{10}$, ${\rm Tr}_{\Gamma(E^0)} \, T - {\rm Tr}_{\Gamma(E^1)} \, T={\rm Tr}_{{\rm Ker}\, D_{10}} T - {\rm Tr}_{{\rm Coker}\, D_{10}} T$. 
Let us decompose the spaces ${\rm Ker}\, D_{10}$ and ${\rm Coker}\, D_{10}$ into subspaces such that each subspace has different eigenvalue for $T$.
If each subspace is finite dimensional, ${\rm Tr}_{{\rm Ker}\, D_{10}} T - {\rm Tr}_{{\rm Coker}\, D_{10}} T$ is well-defined and the following formula holds
\be
{\rm Tr}_{{\rm Ker}\, D_{10}} T - {\rm Tr}_{{\rm Coker}\, D_{10}} T
 = \sum_{x\in {\rm fixed\, point\, set}} \frac{
{\rm Tr}_{E^0_x} \gamma-{\rm Tr}_{E^1_x} \gamma
}{\ |{\rm det}_{T_x X}(1-df(x))|\ }\, .
\ee



\newpage


\appendix



\end{document}